\documentclass[twocolumn]{aastex631} 
\usepackage{savesym}
\usepackage{tipa} 
\savesymbol{tablenum}
\usepackage{siunitx}
\usepackage{comment}
\restoresymbol{SIX}{tablenum}
\DeclareSIUnit{\jansky}{Jy}
\DeclareSIUnit{\MSPS}{MSPS}
\DeclareSIUnit{\byte}{B}
\DeclareSIUnit{\tecu}{TECu}
\DeclareSIUnit{\bit}{bit}
\DeclareSIUnit{\sample}{S}
\DeclareSIUnit{\dmunit}{pc~cm^{-3}}
\DeclareSIUnit{\millisec}{ms}
\newcommand{\kkoname}{k'ni\textipa{P}atn k'l$\left._\mathrm{\smile}\right.$stk'masqt}

\usepackage{cancel}
\usepackage{aas_macros}
\usepackage{ulem}

\usepackage{amsmath, amssymb,commath,bm}
\usepackage{xcolor}
\usepackage{comment}

\DeclareSIUnit{\parsec}{pc}
\newcommand{\todo}[1]{{\color{red}$\blacksquare$~\textsf{[TODO: #1]}}}

\newcommand{\kwm}[1]{{\color{purple}#1}}
\newcommand{\jmp}[1]{{\color{black}#1}}

\newcommand{\finalfix}[1]{{\color{black}#1}}
\newcommand{\fix}[1]{{\color{black}#1}}
\newcommand{\secondfix}[1]{{\color{black}#1}}

\newcommand{\referee}[1]{{\color{black}#1}}
\newcommand{\refereejmp}[1]{{\color{black}#1}}

\newcommand{\matr}[1]{\mathbf{#1}}
\newcommand{\vect}[1]{\bm{#1}}

\received{\today}
\revised{\today}
\submitjournal{AJ}
\shorttitle{A spatial filter for mitigating radio interference}
\shortauthors{Andrew et al.}

\begin{document}
\title{A spatial filter for mitigating \secondfix{radio} interference and its application to CHIME/FRB Outriggers}
\author[0000-0002-3980-815X]{Shion Andrew}
\affiliation{Massachusetts Institute of Technology, 77 Massachusetts Ave, Cambridge, MA 02139, USA}
\affiliation{Department of Physics, Massachusetts Institute of Technology, 77 Massachusetts Ave, Cambridge, MA 02139, USA}

\author[0000-0002-0772-9326]{Juan Mena-Parra}
  \affiliation{Dunlap Institute for Astronomy \& Astrophysics, University of Toronto, 50 St.~George Street, Toronto, ON M5S 3H4, Canada}
  \affiliation{David A.~Dunlap Department of Astronomy \& Astrophysics, University of Toronto, 50 St.~George Street, Toronto, ON M5S 3H4, Canada}

\author[0000-0002-1491-3738]{Haochen Wang}
\affiliation{Massachusetts Institute of Technology, 77 Massachusetts Ave, Cambridge, MA 02139, USA}
\affiliation{Department of Physics, Massachusetts Institute of Technology, 77 Massachusetts Ave, Cambridge, MA 02139, USA}

\author[0000-0002-2510-3124]{Antonios Argyriou}
\affiliation{Department of Electrical and Computer Engineering, University of Thessaly, Volos, 38221, Greece}

\author[0000-0002-4279-6946]{Kiyoshi W. Masui}
\affiliation{Massachusetts Institute of Technology, 77 Massachusetts Ave, Cambridge, MA 02139, USA}
\affiliation{Department of Physics, Massachusetts Institute of Technology, 77 Massachusetts Ave, Cambridge, MA 02139, USA}

  \author[0000-0001-5908-3152]{Bridget C. Andersen}
\affiliation{University of California - Santa Cruz, 1156 High Street, Santa Cruz, CA 95064, USA}
  \author[0000-0003-3772-2798]{Kevin Bandura}
  \affiliation{Lane Department of Computer Science and Electrical Engineering, 1220 Evansdale Drive, PO Box 6109, Morgantown, WV 26506, USA}
  \affiliation{Center for Gravitational Waves and Cosmology, West Virginia University, Chestnut Ridge Research Building, Morgantown, WV 26505, USA}

\author[0000-0001-7166-6422]{Matt Dobbs}
  \affiliation{Department of Physics, McGill University, 3600 rue University, Montréal, QC H3A 2T8, Canada}
  \affiliation{Trottier Space Institute, McGill University, 3550 rue University, Montréal, QC H3A 2A7, Canada}

\author[0000-0001-6128-3735]{Nina V. Gusinskaia}
\affiliation{Anton Pannekoek Institute for Astronomy, University of Amsterdam, Science Park 904, NL-1098 XH Amsterdam, the Netherlands}
\affiliation{ASTRON, the Netherlands Institute for Radio Astronomy, Postbus 2, NL-7990 AA Dwingeloo, the Netherlands}

  \author[0009-0004-4176-0062]{Afrokk Khan}
  \affiliation{Department of Physics, McGill University, 3600 rue University, Montréal, QC H3A 2T8, Canada}
  \affiliation{Trottier Space Institute, McGill University, 3550 rue University, Montréal, QC H3A 2A7, Canada}

\author[0000-0003-2116-3573]{Adam E. Lanman}
\affiliation{MIT Kavli Institute for Astrophysics and Space Research, Massachusetts Institute of Technology, 77 Massachusetts Avenue, Cambridge, MA 02139, USA}
\affiliation{Department of Physics, Massachusetts Institute of Technology, 77 Massachusetts Avenue, Cambridge, MA 02139, USA}

  \author[0000-0002-5857-4264]{Mattias Lazda}
  \affiliation{Dunlap Institute for Astronomy \& Astrophysics, University of Toronto, 50 St.~George Street, Toronto, ON M5S 3H4, Canada}
  \affiliation{David A.~Dunlap Department of Astronomy \& Astrophysics, University of Toronto, 50 St.~George Street, Toronto, ON M5S 3H4, Canada}
  
  \author[0000-0002-4209-7408]{Calvin Leung}
\affiliation{Miller Institute for Basic Research, University of California Berkeley, Berkeley, CA, 94720, USA}
\affiliation{Department of Astronomy, University of California Berkeley, Berkeley, CA 94720, USA}
    
\author[0000-0003-0510-0740]{Kenzie Nimmo}
\altaffiliation{NASA Hubble Fellow.}
\affiliation{Center for Interdisciplinary Exploration and Research in Astronomy, Northwestern University, 1800 Sherman Avenue, Evanston, IL  60201, USA}

\author[0000-0003-0073-5528]{Robert Pascua}
\affiliation{Dunlap Institute for Astronomy and Astrophysics, University of Toronto
50 St. George Street
Toronto, ON M5S 3H4, Canada}
\affiliation{David A.~Dunlap Department of Astronomy \& Astrophysics, University of Toronto, 50 St.~George Street, Toronto, ON M5S 3H4, Canada}
\affiliation{Perimeter Institute for Theoretical Physics
31 Caroline Street North
Waterloo, ON N2L 2Y5, Canada}

\author[0000-0002-8912-0732]{Aaron~B.~Pearlman}
\altaffiliation{NASA Hubble Fellow.}
\affiliation{MIT Kavli Institute for Astrophysics and Space Research, Massachusetts Institute of Technology, 77 Massachusetts Avenue, Cambridge, MA 02139, USA}
\affiliation{Department of Physics, Massachusetts Institute of Technology, 77 Massachusetts Avenue, Cambridge, MA 02139, USA}
\affiliation{Department of Physics, McGill University, 3600 rue University, Montr\'eal, QC H3A 2T8, Canada}
\affiliation{Trottier Space Institute, McGill University, 3550 rue University, Montr\'eal, QC H3A 2A7, Canada}

  \author[0000-0002-3430-7671]{Alexander W. Pollak}
  \affiliation{SETI Institute, 339 Bernardo Avenue, Suite 200 Mountain View, CA 94043, USA}
  
  \author[0009-0005-6633-3945]{Gurman Sachdeva}
  \affiliation{Dunlap Institute for Astronomy \& Astrophysics, University of Toronto, 50 St.~George Street, Toronto, ON M5S 3H4, Canada}
  \affiliation{David A.~Dunlap Department of Astronomy \& Astrophysics, University of Toronto, 50 St.~George Street, Toronto, ON M5S 3H4, Canada}

\author[0000-0002-2088-3125]{Kendrick Smith}
\affiliation{Perimeter Institute for Theoretical Physics
31 Caroline Street North
Waterloo, ON N2L 2Y5, Canada}

\correspondingauthor{Shion Andrew}
\email{shiona@mit.edu}
\collaboration{99}{(CHIME/FRB Collaboration)}







\begin{abstract} 
The sensitivity of radio telescopes is becoming increasingly limited by RFI, which will worsen as the radio spectrum becomes more crowded. One context where this poses a challenge is fast radio burst (FRB) science, where there is increasing scientific interest in capturing as many bursts as possible and accurately measuring their celestial coordinates using interferometry. With several large FRB surveys underway that will be transformative to the field, properly mitigating unwanted interference is essential for the science goals of these surveys to be met. In this work, we present variations of a spatial filter based on the Karhunen–Loève Transform to enhance the sensitivity of interferometers and demonstrate its applicability to FRB detection and localization. We derive a particular variation of the filter for the case of point-like radio pulses, which we show reduces to the maximum-signal-to-noise beamformer. We apply this filter to CHIME/FRB baseband data and demonstrate its capability to enhance the sensitivity and overall localization rate of CHIME/FRB Outriggers. We compare the cross-correlation signal-to-noise obtained using the spatial filter with that obtained using a spectral-kurtosis RFI flagger for a sample of 100 FRBs recorded by CHIME and its Outriggers, and show that this filter will double the total number of FRBs successfully localized with the CHIME/FRB Outrigger telescopes. While demonstrated here in the context of CHIME/FRB Outriggers, the spatial filter presented in this work is broadly applicable to other interferometric radio facilities engaged in FRB science and transient detection, including next-generation telescopes such as CHORD, DSA-2000, BURSTT, and CHARTS.
\end{abstract}

\section{Introduction}
Modern radio telescopes are continually pushing the limits of sensitivity in an effort to detect fainter and more distant sources. However, in practice the sensitivity of radio observations is often limited by sources of radio-frequency interference (RFI), which in many cases can be several orders of magnitude brighter than the desired astrophysical signal \citep{Offringa_2013, CMB_2022, Chen_2025}. With RFI becoming increasingly prevalent and the radio band becoming more crowded, simultaneous progress in the development of RFI mitigation techniques is needed. 

There are many different levels at which RFI mitigation strategies are employed, from physical measures at the telescope site \citep{Raybole_2019} to post-processed data analysis techniques. The latter typically involves some kind of data flagging---first detecting then removing RFI in time, frequency, and/or spatial domains \fix{\citep[see e.g.][]{Deller_2010,Offringa_2012, taylor_2019, Masoud_2023}}. It is important to stress that the efficacy of different RFI mitigation algorithms is highly context specific, depending on properties of the radio telescope being used, the astrophysical signal of interest, and the local RFI environment.

One particularly powerful RFI mitigation strategy well-suited to radio interferometers is spatial filtering of RFI from the desired signal through adaptive beamforming \citep{van_der_veen_2004}. This technique is adaptive because the beamforming \secondfix{weights---which determine the effective beam response---are derived from the measured spatial covariance of the RFI across the array}. Since no a-priori knowledge about the RFI source is required, this technique is particularly useful for types of RFI that are unpredictable in time and frequency. It can also be useful when very faint signals are degraded by persistent, low-level and broadband RFI that cannot be temporally or spectrally excised from the desired signal with traditional flagging methods. Since spatial filters  involve ``spatially excising" out RFI, they perform better the more separable the signal of interest is relative to the unwanted interference in antenna space. In addition to depending on the array geometry and total number of antennas, the separability will depend on the spatial signatures of both the interference and the signal. For instance, point-like sources produce well-defined and highly coherent spatial responses, in contrast to sources that are extended or diffuse. When both the signal of interest and the unwanted interference are compact and originate from different directions, their spatial modes are unlikely to have strong overlap.


There are several reasons \finalfix{why} spatial filters are particularly relevant for studying fast radio bursts (FRBs). FRBs are extragalactic radio transients of unknown astrophysical origins, and because there is a growing scientific interest in localizing as many FRBs as possible to their host galaxies, several next-generation interferometers are being commissioned to capture and precisely localize a larger population of bursts \citep{Hallinan_2019,Vanderlinde_2019,BURSTT_2022,Outriggers_2025,2025_CHARTS}. Both the point-like and transient nature of FRBs, as well as the large interferometric arrays that are being constructed to observe them, align naturally with the capabilities of spatial filters, which can adaptively concentrate and suppress sensitivity in different directions. While spatial filtering can be significantly complicated in contexts such as imaging by its capacity to distort beam shape \citep{Ben_david_2008}, this is irrelevant when studying point sources fully described by a single spatial mode. Moreover, the problem of ``subspace smearing" \citep{Hellbourg_2018}---which arises when the source of interference moves significantly relative to the telescope over the time of the target observation and is therefore spatially ``smeared"---is also significantly mitigated in the context of FRB science where pulses are observed on $\lesssim 1$s timescales.

\fix{For very-long-baseline-interferometry (VLBI) networks, differing RFI environments at each site \secondfix{lead to even larger data loss because cross-correlation is only possible in spectral windows that are free of RFI at all sites}. Consequently, for ``core-outrigger" type networks composed of a central interferometric detector outfitted with smaller stand-alone interferometers---such as the Canadian Hydrogen Intensity Mapping Experiment / Fast Radio Burst  (CHIME/FRB) Outriggers \citep{Outriggers_2025}---fast radio bursts detected by the core can be lost in VLBI cross-correlations due to RFI contaminating additional channels at the outrigger stations. As one of the leading survey-oriented experiments aiming to maximize the total number of precisely localized FRBs, where flux-limited sensitivity directly constrains the accessible survey volume, CHIME/FRB Outriggers will depend heavily on RFI mitigation strategies to fully realize its science goals. }

In this work, we demonstrate the performance of a Karhunen–Loève Transform (KLT) based spatial filter on improving the signal-to-noise of radio pulses in baseband data from the CHIME/FRB instrument. This work is particularly interested in mitigating the number of astrophysical transient sources lost to RFI \fix{\citep[for the related issue of ``false positives", which are particularly relevant in real time backend searches for transients, see][]{Masoud_2023}}. We demonstrate that one of the central advantages of applying this filter in CHIME/FRB baseband data is recovering pulses that were \secondfix{previously unusable for VLBI}. 

In Section~\ref{sec:sky_model}, we develop a formalism to describe the response of an interferometer to a polarized radio sky. In Section~\ref{sec:bf} we present the KL beamformer in the context of signal-to-noise optimization and demonstrate its application for optimal filtering of point source emission. We compare the results of the conventional beamformer to the KL beamformer on real CHIME data in Section~\ref{sec:real_data}, and demonstrate its utility within the context of very long baseline interferometry with CHIME/FRB Outriggers. Caveats and future work are discussed in Section~\ref{sec:conclusion}. 

\section{Sky model and instrument response}
\label{sec:sky_model}

\begin{table*}[t]
    \centering
    \begin{tabular}{l l l }
\hline 
   Symbol & Description & Range/Dimension \\
\hline 
$i,j$ & antenna index & 0 to (2)N \\
$\vect{a}_{i}(\vect{\hat{n}})$ & $i$-th antenna response towards direction $\vect{\hat{n}}$ & vector of size 2 \\
$\vect{d}$ & baseband voltage data & vector of size $(2)N$ \\
$\matr{D}$ & Covariance matrix of the measured voltage data & (2)N X (2)N matrix  \\
$\matr{S}$ & `` " desired voltage signal &`` "  \\
$\matr{R}$ &`` " unwanted interference &`` "  \\
$\matr{N}$ &`` " receiver noise &`` "  \\
$\matr{F}$ & `` " total noise &`` " \\
\hline 
\end{tabular}
\caption{Summary of frequently used notation in this work. Note that $i,j$ index from 0 to N in the case where polarization hands are processed independently (Section \ref{sec:bf}), and index from 0 to 2N  for the more general case when the target's polarization state can be modeled (Appendix \ref{sec:polarization}). }
\label{tab:notation_summary}
\end{table*}

First, we introduce the notation used to describe the data recorded and how it transforms under different variations of our KL filter. Our description of the polarized radio sky and the response of an interferometric array follows the formalism presented in \cite{Shaw_2015} and \cite{Masui_2017}. For convenience, we provide a summary of the indices and symbols used throughout this work in Table~\ref{tab:notation_summary}.

To start, we consider the voltage measured by each antenna in the interferometric array.  The measured voltage (also known as baseband data) of the $i$-th antenna at frequency $\nu$ and time $t$ is given by
\begin{equation}
\label{eq:voltage}
    d_{i}(\nu,t) = \int  \vect{a}_i^\dagger(\nu,t,\vect{\hat{n}}) \vect{\varepsilon}(\nu,t,\vect{\hat{n}}) e^{i2\pi \vect{u}_i^T\vect{\hat{n}}} d\vect{\hat{n}}^2 + n_i(\nu,t)
\end{equation}
where the first term contains the sum of signals over all angles defined by the unit vector $\vect{\hat{n}}$ and the second term contains the receiver noise $n_i$. The complex column vector $\vect{a}_{i} = \left[ A_i^{X}, A_i^{Y}\right]^\dagger$ is the primary voltage beam of the antenna, representing the antenna response to the two polarizations (labeled $X$ and $Y$) of the electric field density vector $\vect{\varepsilon}$. The vector $\vect{u}_{i}=\vect{x}_{i}/\lambda$ is the antenna position in wavelengths. Typically, the antennas of modern interferometers---including CHIME---are dual-polarized, containing pairs of co-located antenna elements with nearly orthogonal polarization response (e.g. $A_i^Y=0$ for an $X$ polarized feed in the ideal case). Nevertheless, since the analog and digital signal processing of the two outputs of a dual-polarization antenna are implemented separately, we treat them as independent elements with different antenna indices $i$. 

Note that in Equation~\ref{eq:voltage}, we made explicit both the frequency and time dependence of these quantities. Since in this work all frequency channels are treated independently, we henceforth omit the $\nu$ and $t$ indices unless it is necessary for clarity of the analysis.

The baseband voltage data $\vect{d}$ allows the estimation of the \jmp{visibilities, the} covariance matrix of the signals from all the antennas in the array, given by

\begin{equation}
\label{eq:data_covariance}
    \matr{D} = \langle \vect{d} \vect{d}^\dagger \rangle = \matr{V} + \matr{N}
\end{equation}
where $\matr{N}=\langle \vect{n} \vect{n}^\dagger \rangle$ is the noise covariance matrix. Under the assumption that the sky is spatially and spectrally incoherent (which is typically the case at radio frequencies), the elements of the \jmp{sky covariance} $\matr{V}$ are given by 
\begin{equation}
\label{eq:vis}
    V_{ij} =\int  \vect{a}_{i}^\dagger \fix{\matr{C}} \vect{a}_{j} ~e^{i2\pi \vect{u}_{ij}^T\vect{\hat{n}}} d\vect{\hat{n}}^2
\end{equation}
where $\vect{u}_{ij}=\vect{u}_{i}-\vect{u}_{j}$ is the baseline vector between antennas $i$ and $j$ in units of wavelength. All the information about the radio sky is captured in \jmp{$\matr{V}$} when the emission is Gaussian distributed. The coherence matrix \fix{$\matr{C}$ is} 
\begin{equation}
\label{eq:coherence_matrix} 
    \matr{C} = \frac{1}{2} \begin{bmatrix}
T+Q & U-iV\\
U+iV & T-Q
\end{bmatrix}.
\end{equation}
The elements of \fix{$\matr{C}$} depend on the Stokes parameters that describe the total intensity ($T$\footnote{We use $T$ to denote total intensity instead of $I$ to avoid potential confusion with the identity matrix $\matr{I}$}), linear polarization ($Q,~U$), and circular polarization ($V$) of the sky emission. The Stokes parameters (and thus \fix{$\matr{C}$}) are functions of frequency and sky position.

In the design of filtering techniques for mitigating interference we will consider the case where the sky covariance matrix $\matr{V}$ consists of some target astrophysical signal, $\matr{S}$, as well as sources of unwanted interference $\matr{R}$---which can be RFI as well as unwanted (but astrophysical) sky background:

\begin{equation}
    \matr{V} = \matr{S} + \matr{R}
\end{equation}
where we have assumed the desired signal to be uncorrelated with the unwanted interference (as well as with the noise $\matr{N}$). This provides us with the following representation of the data covariance:
\begin{equation}
    \label{eqn:data_visibilities}
    \matr{D} = \matr{S} + \matr{F}
\end{equation}
where we have combined the receiver noise $\matr{N}$ and unwanted interference $\matr{R}$ into one noise term  $\matr{F}$. We assume the noise matrix has no noiseless modes (e.g., no faulty receivers) such that $\matr{N}$ (and thus $\matr{F}$) is positive definite.

Finally, for interference mitigation in FRB applications we are interested in the case where the target consists of a single point source (the FRB signal) at location $\vect{\hat{n}}_s$. In this case the elements of the signal covariance $\matr{S}$ simplify to:

\begin{equation}
\label{eq:vis_ps}
    S_{ij} =\vect{a}_{i}^\dagger (\vect{\hat{n}}_s)\matr{C}_{\refereejmp{s}}(\vect{\hat{n}}_s) \vect{a}_{j} (\vect{\hat{n}}_s)~e^{i2\pi \vect{u}_{ij}^T\vect{\hat{n}}_s}.
\end{equation}

\section{Beamforming Optimization}
\label{sec:bf}

\subsection{Single mode analysis}
\label{sec:single_mode}

We now consider the case when the target point source is unpolarized and the signal coherence matrix simplifies to $\fix{\matr{C}_s}\refereejmp{(\vect{\hat{n}}_s)} = \sigma_s^2\matr{I}$, with $\sigma_s^2 = T_s/2$ as the signal power measured by each polarization. We focus on all the antenna elements physically oriented to be mostly sensitive to one of the polarizations, say $X$. For any real array, these elements will also have some spurious sensitivity to the $Y$ polarization. If the array consists of $N$ dual-polarization antennas such that there are $N$ of these particular antenna elements (and $2N$ total antenna elements), the (single-polarization) signal covariance elements can be written as
\begin{align}
\label{eq:S_singlepol}
    S_{ij} = &N\sigma_s^2 \left[A_i^X(\vect{\hat{n}}_s)A_j^{X*}(\vect{\hat{n}}_s)+A_i^Y(\vect{\hat{n}}_s)A_j^{Y*}(\vect{\hat{n}}_s)\right]\times \\\nonumber
    &\frac{e^{i2\pi \vect{u}_{i}^T\vect{\hat{n}}_s}}{\sqrt{N}} \times \frac{e^{-i2\pi \vect{u}_{j}^T\vect{\hat{n}}_s}}{\sqrt{N}}.
\end{align}
Focusing on the term in brackets in Equation~\ref{eq:S_singlepol}, we find 3 cases that make $\matr{S}$ particularly simple:
\begin{enumerate}
    \item The antenna beams are known: In this case the terms in the square brackets are known and can be calibrated out from each signal visibility. 
    \item The antenna beams are identical: In this case the term in brackets becomes a constant for all the signal visibilities and equal to $|\vect{a} (\vect{\hat{n}}_s)|^2$. Since the performance of the KL filter is independent of overall scalings (shown in Sections~\ref{sec:klt_bf} and \ref{sec:klt}), this is equivalent to case 1 above.
    \item The polarization leakage of the antennas is negligible: for our example focused on the $X$ polarization this means that the term in square brackets reduces to $A_i^X(\vect{\hat{n}}_s)A_j^{X*}(\vect{\hat{n}}_s)$ which is factorizable in $i$ and $j$.
\end{enumerate}

In cases 1 and 2, the signal covariance becomes
\begin{align}
\label{eq:S_singlepol1}
    \matr{S} \propto \  &N\sigma_s^2 \vect{\hat{v}}_{s}\vect{\hat{v}}_{s}^\dagger, \hspace{0.2in} v_{s,i}=\frac{e^{i2\pi \vect{u}_{i}^T\vect{\hat{n}}_s}}{\sqrt{N}}
\end{align}
which is a rank-1 covariance matrix, meaning that the signal for this polarization consists of a single sky mode. Case 3 also leads to a rank-1 signal covariance matrix with the change $\vect{\hat{v}}_{s} \rightarrow \matr{Z}\vect{\hat{v}}_{s}$, where $\matr{Z}=\text{diag}\left(A_0^X(\vect{\hat{n}}_s),..., A_{N-1}^X(\vect{\hat{n}}_s) \right)$. Other less used constraints or approximations (such as when the polarization leakage is proportional to the main polarization beam) also lead to a rank-1 signal covariance. In the more general case of the square bracket term in \refereejmp{Equation~\ref{eq:S_singlepol}} the $S_{ij}$ are not factorizable and the  signal covariance will not necessarily be rank-1 (\jmp{but at most rank-2}). This case can be treated using the methods of Section~\ref{sec:klt}.

\subsubsection{Signal power optimization and conventional beamforming}
\label{sec:conv_bf}
Beamforming consists of taking a linear combination of the baseband voltages to enhance the reception of signals from a particular region of the sky\footnote{In this work we focus on pre-correlation beamformers where the beamforming weights are applied to the voltage signals. The treatment and implementation of more general visibility beamformers is discussed in the Appendix, but will be more thoroughly addressed in future works.}. For a beamforming weight vector $\vect{b}$, the beamformed signal is $\vect{b}^\dagger \vect{d}$ and the measured power is

\begin{align}
\label{eq:bf_power}
    \left\langle |\vect{b}^\dagger \vect{d}|^2 \right\rangle  = \left\langle \text{Tr} \left\{ \vect{b}^\dagger  \vect{d} \vect{d}^\dagger \vect{b}\right \} \right \rangle = \vect{b}^\dagger \matr{D} \vect{b}.
\end{align}
Thus, maximizing signal power is equivalent to solving the optimization problem

\begin{align}
\label{eq:bf_opt}
    \underset{\vect{b}}{\text{Max}}\left\{\vect{b}^\dagger \matr{S} \vect{b}\right\}, \hspace{0.2in} \text{subject to}~~\vect{b}^\dagger \vect{b} = 1
\end{align}
where the constraint ensures that we cannot arbitrarily increase power by increasing $|\vect{b}|^2$. The corresponding Lagrangian is

\begin{align}
\label{eq:bf_lagrangian}
    \mathcal{L}(\vect{b}, \lambda)=\vect{b}^\dagger \matr{S} \vect{b}-\lambda\left(\vect{b}^\dagger \vect{b} - 1\right )
\end{align}
where $\lambda$ is the (real-valued) Lagrange multiplier. Setting the derivative of $\mathcal{L}$ with respect to $\vect{b}$ to zero gives

\begin{align}
\label{eq:bf_single_eval_problem}
     \matr{S} \vect{b}=\lambda\vect{b}
\end{align}
which is an eigenvalue problem for $\matr{S}$. Since this is a maximization problem, the optimal $\vect{b}$ is the unit eigenvector corresponding to the largest eigenvalue of $\matr{S}$. For the particular case of Equation~\ref{eq:S_singlepol1}, the optimal beamforming kernel is $\vect{b}=\vect{\hat{v}}_{s}$ and the optimized signal power is
\begin{align}
\label{eq:bf_power_opt}
     \vect{\hat{v}}_{s}^\dagger \matr{S} \vect{\hat{v}}_{s} = \lambda = N\sigma_s^2.
\end{align}
This delay-and-sum beamformer, also referred to as the ``conventional beamformer", is the most common and widely used beamforming algorithm due its simplicity \citep{michilli2020analysis}. Other beamforming weights can be used if, for example, we prioritize having more control over the shape of the synthesized beam \citep{Masui_2017}.

For case 3 given in Section \ref{sec:single_mode}, the beamformer and respective measured power become 

\begin{align}
\label{eq:bf_power_opt_case3}
    \vect{b}&=\frac{\matr{Z}\vect{\hat{v}}_{s}}{(\vect{\hat{v}}_{s}^\dagger \matr{Z}^\dagger\matr{Z} \vect{\hat{v}}_{s})^{1/2}} = \frac{N^{1/2}\matr{Z}\vect{\hat{v}}_{s}}{\left [\sum_i |A_i^X(\vect{\hat{n}}_s)|^2\right]^{1/2}} \\
     \vect{b}^\dagger \matr{S} \vect{b} &=  N\sigma_s^2~\vect{\hat{v}}_{s}^\dagger \matr{Z}^\dagger\matr{Z} \vect{\hat{v}}_{s}= \sigma_s^2\sum_i |A_i^X(\vect{\hat{n}}_s)|^2.
\end{align}
\subsubsection{Signal-to-Noise optimization and KL beamforming}
\label{sec:klt_bf}

As shown in Section~\ref{sec:conv_bf}, the pointed beams of the conventional beamformer maximize signal power in a particular direction of the sky. However, as explained in \cite{Masui_2017}, this beamformer is not necessarily optimal in that it does not necessarily maximize the signal-to-noise ratio, $S/N$,  given by
\begin{align}
\label{eq:snr_bf}
    S/N = \frac{\vect{b}^\dagger \matr{S} \vect{b}}{\vect{b}^\dagger \matr{F} \vect{b}}.
\end{align}
In the CHIME/FRB real-time transient search backend, it is $S/N$, and not signal power, that is used to determine whether a transient event becomes an FRB candidate and whether baseband data are saved to disk for further analysis. Likewise, in CHIME/FRB Outriggers, $S/N$, rather than signal power, is used to assess the reliability of VLBI fringes and to determine whether a refined FRB localization is possible. 

A non-trivial $\matr{F}$ ($\mathbf{F}\not\propto \mathbf{I}$) can arise from noise couplings between the antennas of the array or from unwanted sky signals---including RFI. In those cases we can optimize the beamforming weights to enhance signal power while controlling the effect of these unwanted signals on the measured noise. The optimization problem becomes

\begin{align}
\label{eq:klt_bf_opt}
    \underset{\vect{b}}{\text{Max}}\left\{\frac{\vect{b}^\dagger \matr{S} \vect{b}}{\vect{b}^\dagger \matr{F} \vect{b}}\right\}.
\end{align}

\jmp{Since multiplying $\vect{b}$ by a scalar leaves the $S/N$ unchanged, we can restate the problem as}

\begin{align}
\label{eq:klt_bf_opt1}
    \underset{\vect{b}}{\text{Max}}\left\{\vect{b}^\dagger \matr{S} \vect{b}\right\}, \hspace{0.2in} \text{subject to}~~\vect{b}^\dagger \matr{F} \vect{b} = 1.
\end{align}
Following the procedure in Section~\ref{sec:conv_bf} leads to
\begin{align}
\label{eq:klt_bf_single_eval_problem}
     \matr{S} \vect{b}=\lambda \matr{F}\vect{b}
\end{align}
which is a generalized eigenvalue problem for $\matr{S}$ and $\matr{F}$, and is also the base for the KLT (also known as the signal-to-noise eigendecomposition).
For the particular case of Equation~\ref{eq:S_singlepol1}, the optimal KL beamformer is 

\begin{align}
\label{eq:klt_bf}
     \vect{b}=\frac{ \matr{F}^{-1}\vect{\hat{v}}_{s}}{\left( \vect{\hat{v}}_{s}^\dagger \matr{F}^{-1}\vect{\hat{v}}_{s}\right)^{1/2}}
\end{align}
and the optimal $S/N$ is
\begin{align}
\label{eq:klt_bf_snr}
     \left( S/N \right)_{KLT} = \lambda = N\sigma_s^2~ \vect{\hat{v}}_{s}^\dagger \matr{F}^{-1}\vect{\hat{v}}_{s}.
\end{align}

Note that when $\matr{S}$ consists of a single mode the KL beamformer is $\vect{b} \propto \matr{F}^{-1}\vect{\hat{v}}_{s}$ and belongs to the family of other well known beamformers including the matched filter \citep{1444313} and the minimum variance distortionless response (MVDR) beamformer \citep{Capon_1969}. These filters have the same basic form, with the normalization constant set by the desired normalization of the synthesized beam. Also note that when $\matr{F}\propto \matr{I}$, Equations~\ref{eq:klt_bf}
and \ref{eq:klt_bf_snr} reduce the respective conventional beamformer expressions. \referee{In general, the relative signal-to-noise improvement between the optimal KL beamformer and the conventional beamformer will depend on  the relative spatial signature of the desired signal relative to the unwanted interference, which we show quantitatively in Appendix~\ref{sec:toy_model} to be larger the more spatially separable the signal and the interference are. However, the KL filter is guaranteed by construction to outperform any other beamformer of the same form regardless of the location of the interference relative to the target as long as the signal and contamination covariances are well-known. }

\subsection{Multiple signal modes, polarization, and the KL transform}
\label{sec:klt}

In Section~\ref{sec:conv_bf} we already identified a case where a single spatial beamformer cannot capture all the information about the point source of interest. Another case relevant for FRBs is when the source is polarized, which we investigate in more detail in Section~\ref{sec:polarization}. In cases when the signal of interest spans several independent modes\footnote{Here we use mode to refer to any linear combination of the data.}, what we need is a set of statistically uncorrelated beamformers that measure the respective signal power while controlling the effect of unwanted contamination. Packing those beamformers as the columns of a matrix $\matr{B}$, 
\jmp{we can extend Equation~\ref{eq:klt_bf_opt1} to include multiple modes by considering the optimization problem}

\begin{align}
\label{eq:klt_opt_multimode}
    \underset{\matr{B}}{\text{Max}}\left\{\text{Tr} \left(\matr{B}^\dagger \matr{S} \matr{B}\right)\right\}, \hspace{0.2in} \text{subject to}~~\matr{B}^\dagger \matr{F} \matr{B} = \matr{I}
\end{align}
\jmp{which} leads to
\begin{align}
\label{eq:klt_eval_problem}
     \matr{S} \matr{B}= \matr{F}\matr{B}\matr{\Lambda}
\end{align}
where $\matr{\Lambda}$ is a real-valued diagonal matrix \citep{2019arXiv190311240G}. This is \jmp{the general form of} the generalized eigenvalue problem for $\matr{S}$ and $\matr{F}$. Its solution gives the KLT matrix $\matr{B}$ which, when applied to the data $\vect{d}$, jointly diagonalizes the signal $\matr{S}$ and contamination $\matr{F}$ covariance matrices. This means that

\begin{align}
\label{eq:klt_transform}
     \vect{\tilde{d}} = \matr{B}^\dagger\vect{d}
\end{align}
leads to 

\begin{align}
\label{eq:klt_diagonalized_cov}
     \langle \vect{\tilde{d}} \vect{\tilde{d}}^\dagger \rangle = \matr{\tilde{D}} = \matr{B}^\dagger (\matr{S}+\matr{F}) \matr{B}=\matr{\Lambda}+\matr{I}.
\end{align}

In this new basis, the generalized eigenvalues in the diagonal of $\matr{\Lambda}$ represent the signal-to-noise power measured in the respective eigenmode, that is, the $S/N$ measured by $\vect{b}_{k}$, the beamformer in the $k$-th column of $\matr{B}$, will be

\begin{align}
\label{eq:klt_snr_k}
     (S/N)_{KLT,k}=\vect{b}_{k}^\dagger \matr{S} \vect{b}_{k} = \lambda_k.
\end{align}
\referee{If we know that the power of the signal of interest is contained within specific modes we can isolate the corresponding signal subspace by forming $\matr{\tilde{B}}$, a matrix whose columns have only the modes of interest.} \jmp{The optimal value of our objective function in Equation~\ref{eq:klt_bf_opt} will be

\begin{align}
\label{eq:klt_snr_tot}
     \underset{\matr{\tilde{B}}^\dagger \matr{F} \matr{\tilde{B}} = \matr{I}_k}{\text{Max}}\left\{\text{Tr} \left(\matr{\tilde{B}}^\dagger \matr{S} \matr{\tilde{B}}\right)\right\} = \sum_{k~\text{a signal  mode}} \lambda_k.
\end{align}}
If the signal is contained in a few modes ($\matr{S}$ not full rank), then the remaining generalized eigenvalues should be zero and adding them should not affect the total $S/N$. In practice, the generalized eigenvalue spectrum will have a steep transition, with a few large values corresponding to signal power, followed by several residual eigenvalues that are small but not zero due to effects such as finite integration times for covariance estimation, rounding errors, etc.

\jmp{The KLT is the closed-form solution to the optimization problem in Equation~\ref{eq:klt_snr_tot}, with well-understood and computationally efficient algorithms available for solving the associated generalized eigenvalue decomposition. However, there is more than one way to extend the single-mode $S/N$ optimization problem in Section~\ref{sec:klt_bf} to multiple modes, and it is not immediately obvious that Equation~\ref{eq:klt_snr_tot} is the most appropriate performance metric to optimize. 

For instance, based on Equations~\ref{eq:snr_bf} and \ref{eq:klt_bf_opt}, one may instead formulate a criterion based on a more direct definition of the total signal power to total noise power ratio for a signal spanning several independent modes:

\begin{align}
\label{eq:trace_ratio_multimode}
     \underset{\matr{\tilde{B}}^\dagger \matr{\tilde{B}} = \matr{I}_k}{\text{Max}}\left\{\frac{\text{Tr} \left(\matr{\tilde{B}}^\dagger \matr{S} \matr{\tilde{B}}\right)}{\text{Tr} \left(\matr{\tilde{B}}^\dagger \matr{F} \matr{\tilde{B}}\right)}\right\}.
\end{align}

When $k=1$ the formulations in Equations~\ref{eq:klt_snr_tot} and ~\ref{eq:trace_ratio_multimode} coincide, and the KLT provides the optimal solution (Section~\ref{sec:klt_bf}). For $k>1$, however, the two formulations optimize different objectives. The trace ratio in Equation~\ref{eq:trace_ratio_multimode} maximizes the total $S/N$ of the $k$-dimensional subspace, whereas Equation~\ref{eq:klt_snr_tot} maximizes the sum of the individual mode $S/N$ values, equivalent to selecting the leading generalized eigenmodes after pre-whitening the data. 

Which approach \referee{is optimal depends on which definition of ``signal-to-noise'' is most appropriate for a} specific application and computational constraints. While conceptually appealing as a direct generalization of the total $S/N$, the trace ratio problem does not admit a closed-form solution in general and is not always solved by the KLT \citep[for detailed reviews on the topic in the context of machine learning and high-dimensional data analysis see][]{doi:10.1137/120864799,FERRANDI2025108108}. Obtaining the true optimum requires iterative algorithms that typically involve solving an eigenvalue problem (or comparably expensive operations) at each iteration, making the approach computationally more demanding for large datasets.

\referee{In this work, we adopt the objective function in Equation~\ref{eq:klt_snr_tot} (the sum of per-mode $S/N$) instead of Equation~\ref{eq:trace_ratio_multimode} (the trace ratio or ``total" S/N of the subspace)} because of its simple implementation via the KLT and because its closed-form solution facilitates the polarization analysis presented in Appendix~\ref{sec:polarization}. It will also facilitate future analyses of the robustness of the multi-mode filter to calibration, pointing, and polarization errors.} Moreover, performing a single generalized eigenvalue decomposition directly yields the $S/N$ associated with each mode, which simplifies the selection of modes with minimal noise contamination when additional compression of the multi-mode data is required. \referee{We also note that for the central application considered in this work–––VLBI with the Outrigger stations–––the per-mode formulation is more appropriate, since individual modes are cross correlated separately (whether they be spatial modes or polarimetric modes) between stations. } 


\subsection{Estimating Covariances}
\label{sec:cov_estimation}
The KL filter requires a model for $\matr{S}$ and an estimate for $\matr{F}$. Since the flux of the source only changes $\matr{S}$ by an overall constant, only the steering vector of the beam  (given by $\vect{\hat{n}}_s$) and polarization state of the source (given by \jmp{$\mathbf{C}$}) need to be specified to model $\matr{S}$ in the case of a point source and a properly calibrated interferometer. The two main sources of error in estimating $\matr{S}$ for a point source that is stationary over the time of observation are gain errors---which arise from imperfect antenna calibration---and pointing errors---which arise from not knowing the true target location ($\vect{\hat{n}}_s$). \referee{In Appendix~\ref{sec:toy_model}, we quantitatively investigate the effect of both these systematics on the performance of the KL beamformer compared to the conventional beamformer, with calibration errors discussed in section ~\ref{subsec:phase_errors} and pointing errors discussed in Section ~\ref{subsec:Pointingerrors}}. 

For frequency channels that are either interference or noise dominated, the eigenvalues characterizing all contaminants will be much larger than those characterizing the target signal, in which case $\matr{F}$ can be obtained from the sample covariance of the data. In the case of transients where the expected arrival time of the astrophysical pulse is already known, $\matr{F}$ can be obtained for \emph{any} frequency channel by computing the sample covariance immediately before or after the pulse (typically $\sim10$\rm{ms} long), under the assumption that $\matr{F}$ is stationary on short timescales.

Finally, we note that since the sample covariance matrix over $M$ time samples is given by 
\begin{equation}
    \label{eq:diagonal_loading}
    \hat{F}_{ij} = \frac{1}{M}\sum_{\jmp{m}} d_i(t_\jmp{m})d_j^*(t_\jmp{m})
\end{equation}
where $\matr{\hat{F}}$ is constructed by $\sim$M outer products (rank-1 matrices), a \emph{minimum} of $N$ time samples are needed to ensure that $\matr{F}$ is full rank, a requirement for constructing the filter.
In principle, ``diagonal loading" ($\matr{F}\rightarrow \matr{F}+\epsilon \matr{I}$ where $\epsilon>0$ is a small scalar) can be used to enforce positive definiteness in $\matr{F}$; however, we find that in our data (typically over $\sim10^5$ time samples) this is generally not required for CHIME or its outrigger stations ($N\sim 10^3$)\referee{, and that the eigenvalue spectrum of the noise covariance does not change appreciably over the course of our observations ($< 1s$)}.

\section{Application to CHIME/FRB data}
\label{sec:real_data}
CHIME and its Outriggers are each stand-alone interferometers operating from 400-800\,MHz. CHIME itself contains 1024 dual-polarized feeds, its \kkoname \  Outrigger (KKO) contains 64 dual-polarized feeds, and its Outriggers in Green Bank (GBO) and Hat Creek (HCO) contain 128 dual-polarized feeds \citep{chime_frb_2018,kko_adam,Outriggers_2025}. Upon detection of a radio transient, CHIME sends triggers to all its Outriggers to record $\sim$100ms of baseband data encompassing the burst. The baseband data contains 1024 spectral channels each with a resolution of 390\,kHz (sampled in the second Nyquist zone at 2.56\,$\mu$s). The real-time CHIME/FRB detection backend, which searches for transients over beams that are fixed across the sky, is able to localize the transient to several arcminutes, which can be refined to sub-arcminute precision with a finer beamforming grid search conducted offline using the recorded baseband data \citep{Michilli_2021}. In either case, a precise enough localization is known such that the signal of interest $S$ can be taken as a well-modeled point source at each individual station  (see Appendix \ref{subsec:Pointingerrors} for discussion on the effect of pointing errors). 

Since the science goals of most radio interferometers require much more precise calibration for interferometry than for polarimetry, modeling the polarization state of $\mathbf{S}$ is generally not nearly as accurate as modeling its interferometric state. \referee{As stated in the end of Section \ref{sec:klt}, due} to the practical difficulties of modeling the full polarization state of $\mathbf{S}$, particularly at the Outrigger telescopes, we perform the filter on each polarization state of CHIME baseband data independently---this is equivalent to the treatment presented in Section \ref{sec:klt_bf}. \referee{However, given that RFI itself is often polarized, we note that investigating the performance improvement seen by the polarimetric form of the filter is an important direction for future work.}

\subsection{Results on a beamformed FRB pulse}
\begin{figure}[h]
    \centering
    \includegraphics[width=\linewidth]{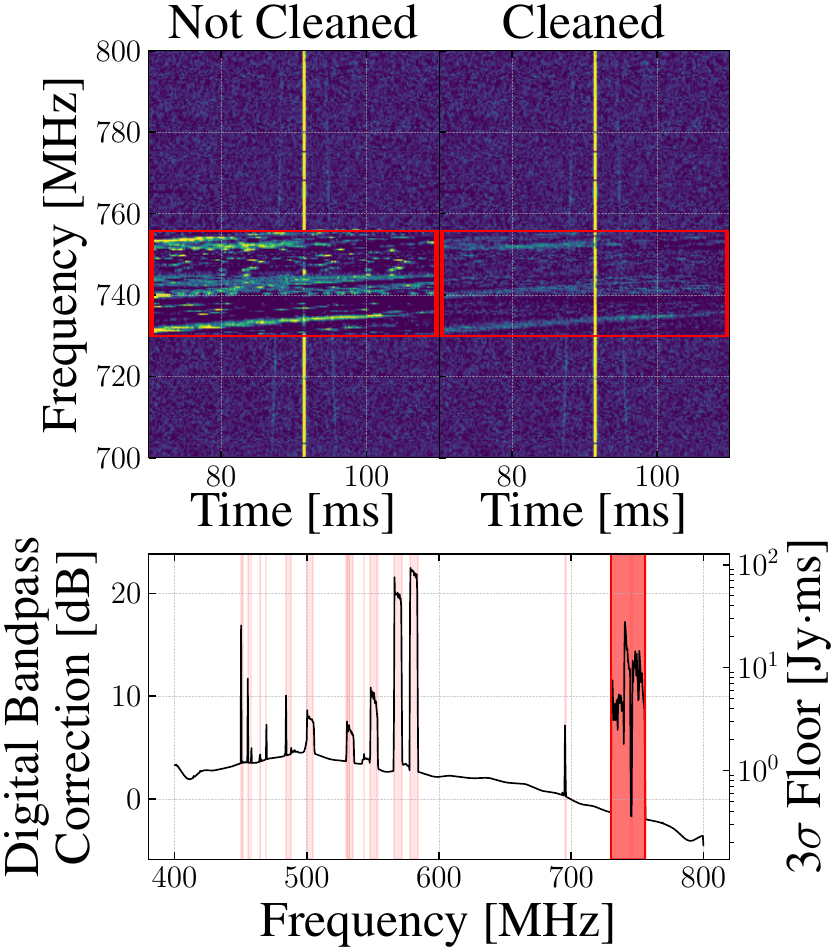}
    \caption{\fix{Top panel: comparison of a beamformed pulse from FRB~20210603A before (left) and after (right) applying the KL filter for the top 100\,MHz of the CHIME band. The RFI contaminated channels in this part of the band are highlighted in red. This demonstrate\refereejmp{s} the filter's ability to recover a pulse in channels strongly contaminated by RFI. Bottom panel: here we denote the inverse of the digital gains---averaged over all receiver and digitizer chains---as the ``digital band\refereejmp{-}pass correction", which are shown in units of dB as a function of frequency. RFI contaminated channels, which are shaded red, are clearly visible from the significant jumps in the digital gains, where the average RFI power levels are over 10\,dB larger than the noise in clean channels. In dark red are the RFI contaminated channels spanning $\sim730-756$\,MHz}.}
    \label{fig:morphology}
\end{figure}

\begin{figure}[h]
    \centering
    \includegraphics[width=\linewidth]{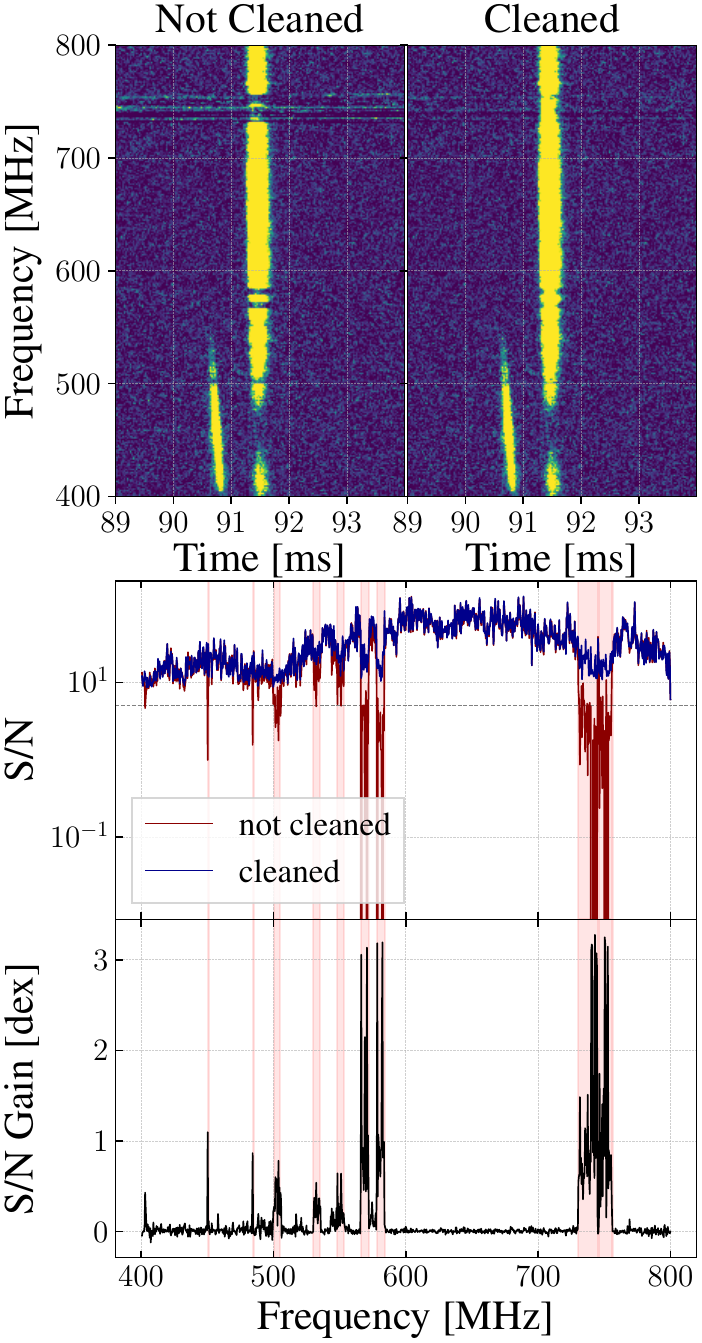}
    \caption{\fix{Top panel: comparison of a beamformed pulse from FRB~20210603A before (left) and after (right) applying the KL filter over the full band. Middle and bottom panels: Improvement in the  signal-to-noise (ratio) of the beamformed pulse (in units of dex) after applying the filter as a function of frequency. Some RFI-contaminated channels (identified in the bottom panel of Figure~\ref{fig:morphology}) are shaded in red, where we see the most notable signal-to-noise improvement.}}  \label{fig:_pulse_full_band.pdf}
\end{figure}

We first demonstrate the performance of our filter on FRB~20210603A and its ability to recover signal from a part of the CHIME band that is heavily RFI-contaminated. This particular FRB was chosen because it is one of the few archival CHIME datasets with recorded baseband data in all frequency channels---in more recent observations, only 80\% of the full band is recorded at CHIME due to a fraction of X-engine nodes being offline. Figure~\ref{fig:morphology} compares a waterfall plot of the FRB from $732-745$\,MHz after beamforming to its known VLBI position \citep{leung2020synoptic,Cassanelli_2024} with and without the KL filter. The RFI in this part of the band is persistent and $\sim$ $10-100$ times stronger than the nominal system noise level, which can be seen from the digital band-pass correction applied to CHIME's analog-to-digital converters plotted in Figure~\ref{fig:_pulse_full_band.pdf}. 

Figure \ref{fig:_pulse_full_band.pdf} also shows a signal-to-noise comparison between the beamformed pulse with and without the filter over the entire band. The spikes in the digital gain correction correspond to persistent RFI, which as expected is also where we see the largest signal-to-noise improvement in the KL filter. Here we make two important notes. The first is 
\referee{that \cite{menaparra_quantization_bias} found that in the case that the quantization levels are configured to be optimal (which is the case for CHIME and its Outriggers), the effect of quantization errors on visibility estimates is uncorrelated additive noise; therefore, although the covariance estimation for the filter should not be biased, the frequency-dependent quantization noise set by the digital gains places a fundamental floor on how bright a source must be to be detected even if a post-processing filter perfectly removes all sources of interference. For example, a 1 Jy$\cdot$ms source could in theory be detected at $3\sigma$ at 780\,MHz at CHIME but could never be detected at 750\,MHz, which is heavily RFI contaminated}. The second is that the signal-to-noise improvement is not limited to just RFI contaminated channels and instead applies to the entire band---the filter, after all, is constructed by design to maximize the signal-to-noise, where the noise can be sourced from sky backgrounds more generally as well as RFI. Such a filter can help constrain astrophysical measurements, such as rotation measure or scintillation, that are strongly frequency dependent; but even more powerful is its ability to recover a pulse that previously was undetectable, which we discuss within the context of VLBI in the following section. 



\begin{figure}[h]
    \centering
    \label{fig:digital_gains}
    \includegraphics[width=\linewidth]{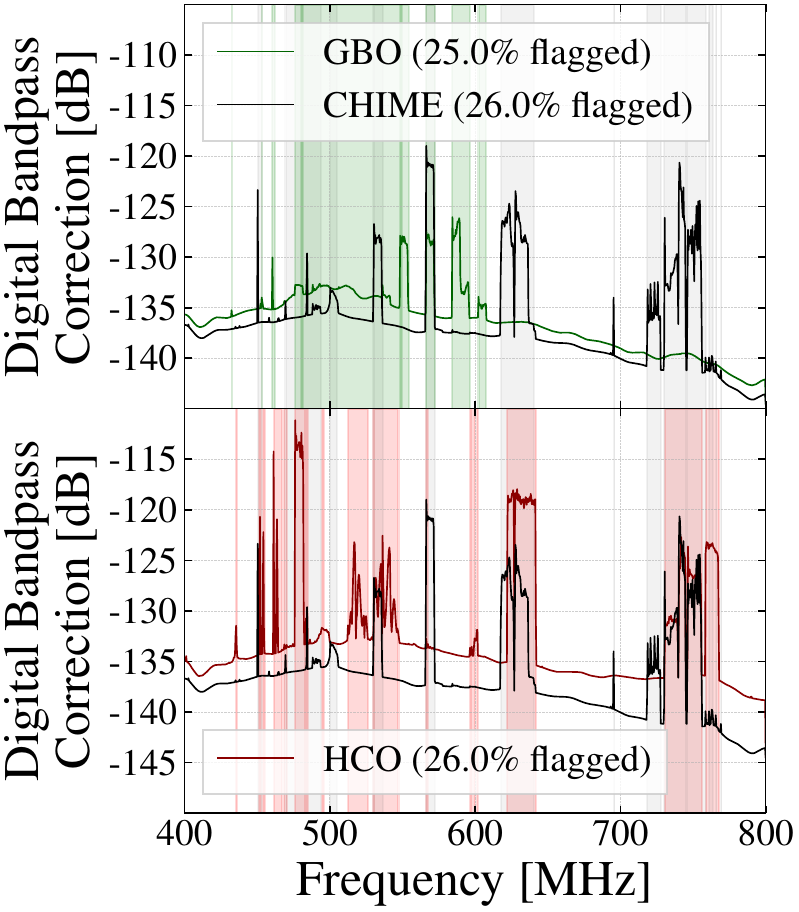}
    \caption{Comparison of the non-overlapping RFI environments at HCO, GBO and CHIME, represented by the inverse of the digital gains averaged over all receiver and digitizer chains. At each individual station, $\sim$ 20-30\% of the band on average is lost due to persistent RFI, and about half of the band is lost to RFI in the CHIME-Outriggers cross-correlated data.}\label{fig:Outrigger_digital_gains}
\end{figure}

\begin{figure}[h]
    \centering
    \includegraphics[width=\linewidth]{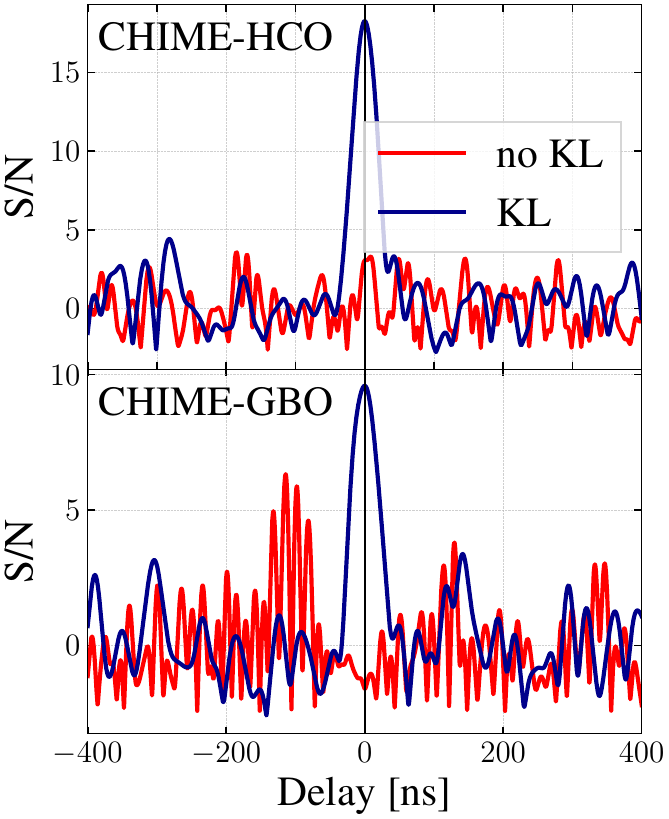}
    \caption{Comparison of cross-correlation signal-to-noise in CHIME-Outrigger cross-correlated VLBI data of pulsar B2154+40, calibrated using J2154+4515. The correct solution \secondfix{corresponding to the known position of the pulsar \citep{Chatterjee_2009}} is at 0\,ns, which is clearly inconsistent with the \referee{RFI}-contaminated fringes in red.}  \label{fig:VLBI}
\end{figure}

\begin{figure}[h]
    \centering
    \includegraphics[width=\linewidth]{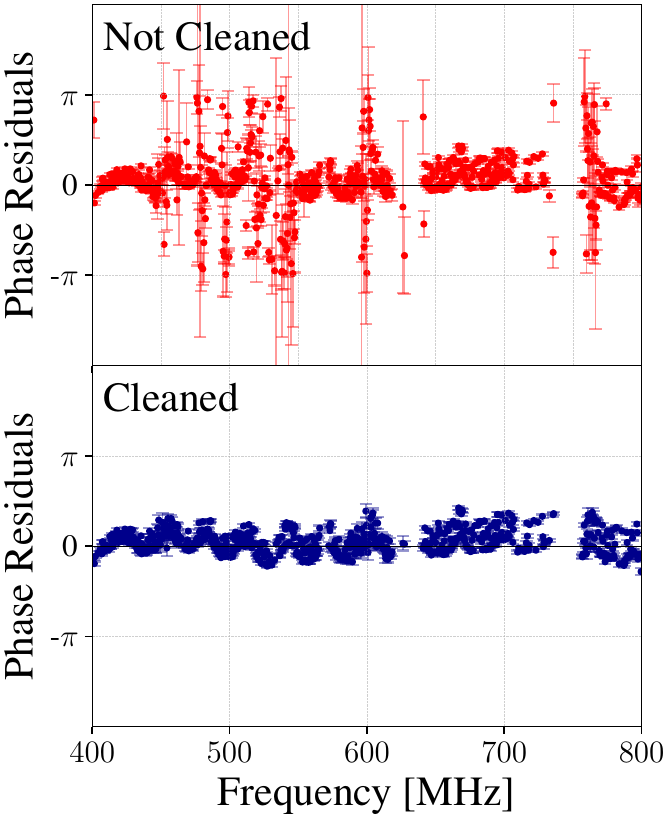}
    \caption{Comparison of phase residuals in CHIME-HCO cross-correlated VLBI data for FRB~20250316A \citep{RBFLOAT_2025ApJ} without the KL filter (top) and with the KL filter (bottom). Note that in both panels, only frequencies not recorded by CHIME itself have been masked out. The residual $\sim$ \refereejmp{30~MHz} ripple is \referee{a known systematic} due to a differential beam phase \citep{kko_adam,Outriggers_2025} and is not an artifact of RFI.}  \label{fig:VLBI_bright}
\end{figure}

\begin{figure}[h]
    \centering
    \includegraphics[width=\linewidth]{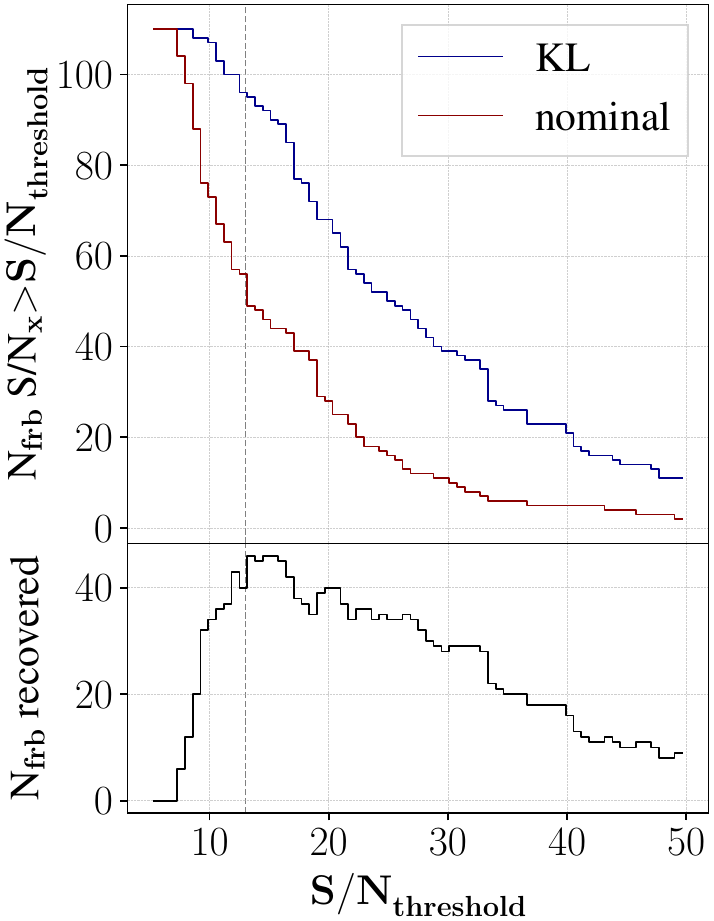}
     \caption{Top panel: Total number of FRBs with a cross-correlation signal-to-noise above the threshold specified. The sample shown is all CHIME-detected FRBs with Outriggers data successfully recorded at all Outrigger sites from February to August 2025. The minimum cross-correlation signal-to-noise ratio required for a robust localization is 13, and indicated by the dashed gray line. Bottom panel: the total number of FRBs recovered after applying the filter with a cross-correlation signal-to-noise ratio above the specified threshold. The fact that the distribution peaks at the cusp of a ``detection" is a function of the intrinsic flux distribution of FRBs and highlights the utility of this filter at maximizing the total localization rate.}  \label{fig:VLBI_SNR_CDF}
\end{figure}

\subsection{CHIME/FRB Outriggers}
The need for a large sample of precisely localized FRBs has motivated a new generation of instruments typically consisting of a ``central core" interferometer and ``outrigger" or ``auxiliary" antennas \citep[e.g.,][]{Vanderlinde_2019,BURSTT_2022,Outriggers_2025}. While outrigger antennas that form long baselines with the central core are needed to obtain the angular resolution necessary for precise localizations, they are generally not included in the real-time FRB backend search---which are compute limited and typically optimized to maximize the field of view/FRB detection rate. Instead, only the central core, which contains most of the collecting area, participates in the FRB search, and upon an FRB detection sends triggers to the outrigger antennas so that the FRB data is recorded and localization can be done offline. 

In the case where the only source of noise is thermal, the signal-to-noise ratio of a correlation function scales with the number of independent pairs as $\sqrt{N_{\rm{pairs}}}$. This means that the signal-to-noise in an autocorrelation at a single station where there are $N_1$ antennas ($N_{\rm{pairs, auto}}=N_1^2/2$) differs from the signal-to-noise in a cross-correlation between two different stations with $N_1$ and $N_2$ antennas by a factor of $\sqrt{2}$ if $N_1=N_2$  since $N_{\rm{pairs, cross}}=N_1N_2$. The fact that, in the ideal case, a factor of $\sqrt{2}$ can be recovered in cross-correlation signal-to-noise is why outrigger stations often contain less than half the collecting area of the central core \citep{Outriggers_2025}. However, in reality the noise environments---and in particular, the RFI environments---can be dramatically different for Outrigger sites that are sufficiently far away from the central core. As such, FRBs with a signal in the central core are not guaranteed to be detectable in cross-correlation, which both reduces the total localization rate and complicates Outrigger selection effects. Moreover, for VLBI networks that rely on a large fractional bandwidth to fit out systematics \citep[see][]{Outriggers_2025}, the localization precision is degraded in the presence of RFI even if fringes are obtained. 

Naively, one might expect RFI to ``correlate out" when cross-correlating baseband data between two telescopes on long ($\gg10\,\rm{km}$) baselines. Even in an idealistic Gaussian interference case, this can only hold if the RFI amplitude is no larger than the desired signal by $\sim \sqrt{M}$---where the integration period and therefore the number of time samples $M$ is not adjustable in the case of transients. \fix{In other words, RFI tends to be too bright and transient events too short-lived for the RFI to sufficiently decorrelate when averaging over a pulse}. This assumption is further exacerbated in the case where the total collecting area (total number of feeds over the entire network) is concentrated at an individual station where RFI \emph{is} expected to correlate (i.e., the total number of baselines \fix{long enough for the RFI to decorrelate} is small), as in the case where each station is a standalone interferometer. For instance, Figure~\ref{fig:Outrigger_digital_gains} shows the non-overlapping nature of RFI environments at CHIME as compared to its Outrigger stations.


\subsubsection{Results on VLBI cross-correlated visibilities}
We test the performance of our filter on a baseband dataset encompassing a burst from the pulsar B2154+40. The KL filter was applied independently to CHIME and each Outrigger site according to the beamforming procedure described in Section \ref{sec:real_data}. In order to mimic the effect of beamformed pointing errors for FRBs with positions unknown at the subarcsecond level, the source model \jmp{$\matr{S}$} was calculated by a pertu\refereejmp{r}bation of $1'$ to the ``true" position based on  astrometric observations from the VLBA \citep{chatterjee2017direct}. Beamformed data with and without the filter were also made for the calibrator J2154+4515 with no pertu\refereejmp{r}bations applied to \jmp{$\matr{S}$} \citep{Andrew_2025}. 
We then correlated all beamformed data with the PyFx correlator \citep{Leung_2025} to form ``KL filtered" and ``unfiltered" visibilities on each baseline.  

Here, our goal for the filter is to improve the measured signal-to-noise of the visibilities, but without degrading our astrometry.  As such, although we filter and beamform to a perturbed target position, we \emph{correlate} to the ``correct" target position, since an incorrect pointing at the beamforming stage will reduce the sensitivity (see Appendix \ref{ap:extensions} for more discussion) but not bias the VLBI astrometry.  We then calibrate our target visibilities to J2154+4515 under the expectation that the measured residual delay in the target is very close to 0. 

The fringing strength of our visibilities on the CHIME-HCO baseline and CHIME-GBO baseline with and without the filter are compared in Figure~\ref{fig:VLBI}. Without the filter, strong fringing around the expected delay is not observed on either baseline. With the filter, a clear signal is recovered at a delay consistent with the expected value of 0\,ns. 

We also test our filter on VLBI visibilities of FRB~20250316A \citep{RBFLOAT_2025ApJ}, which is bright enough to calculate and compare signal-to-noises on a frequency channel-by-channel basis. The phase residuals of the FRB visibilities after calibration \fix{and correlation to the best-fit position} are shown in Figure~\ref{fig:VLBI_bright}, where significantly less scatter is observed in particular in parts of the band that are heavily RFI contaminated. Since the astrometric precision of Outrigger localizations rely strongly on a large fractional bandwidth to remove the differential delay due to the ionosphere in the fringe fit \citep{Outriggers_2025}, Figure~\ref{fig:VLBI_bright} hints at the additional potential of this filter to improve localization accuracy. This will be characterized in detail in an upcoming work dedicated to the astrometric performance of the Outrigger telescopes. 


\subsubsection{ Broader impact on VLBI recovery rate}
The central goal of the CHIME/FRB Outriggers project is to localize a large fraction of CHIME-detected FRBs to within their host galaxies. Based on the collecting areas of the Outrigger cylinders, this in principle should be possible for all FRBs with a CHIME $S/N$ of 10 even after accounting for direction-dependent systematics \citep{Outriggers_2025}, but in practice is complicated by the existence of strong persistent RFI corrupting $\sim20-30\%$ of the band at each telescope site in a largely non-overlapping way. This matters the most for the faintest FRBs with fluxes comparable to the noise floor of the RFI, of which there are also the most bursts at CHIME and its Outrigger stations. 

We highlight the central utility of this filter within the context of CHIME/FRB Outriggers in Figure~\ref{fig:VLBI_SNR_CDF}. In order to obtain a representative sample, we examined data that was obtained over different parts of the year to capture the fluctuations in RFI environments and system temperature that occur on $\sim$monthly timescales. More specifically, we compare the distribution of cross-correlation signal-to-noises for a sample of all the Outrigger-triggered FRBs that were successfully recorded at all stations between February to August 2025. There is a clear boost in signal-to-noise over the entire sample---and critically, the number of ``localizable" FRBs above a minimum reliable signal-to-noise threshold is nearly doubled after applying the filter.







\section{Conclusion}
\label{sec:conclusion}
We have presented a spatial filter for improving the signal-to-noise ratio of interferometric measurements based on the KL Transform. We test our algorithm on CHIME/FRB data, and show that the filter not only improves the CHIME signal-to-noise of an FRB pulse by several orders of magnitude in RFI-contaminated channels, but also consistently improves the signal-to-noise over the full band. We also show that this filter can be applied independently to CHIME and each of its Outrigger stations to improve the signal-to-noise of VLBI cross-correlations. Finally, we demonstrate that one particularly useful application of this filter is maximizing the total localization rate of radio transients for core+outrigger-like telescopes designed similar to CHIME/FRB Outriggers. 

More broadly, we have also outlined a framework in which variations of this filter can be described and tested, as well as the caveats that pointing errors, gain errors, and nonlinearities in the correlator can introduce to the filter's performance. Although we have focused on a particular variation of the filter in this work, the KL formalism presented is flexible enough to be extended to other contexts. For instance, given that RFI is typically polarized, it is likely that this filter can be made significantly more effective with the incorporation of polarimetry as additional discriminating information. Higher-order moments beyond the covariance can also be used to excise sources of interference from the data using visibility beams. The software associated with this work is publicly available on Github\footnote{\url{https://github.com/shionandrew/KLT\_filter\_public}} and includes some of the aforementioned extensions. We hope that some of the framework presented in here can be built on and eventually tested in future work. 

\section{Acknowledgments}

We acknowledge that CHIME and the \kkoname{} Outrigger (KKO) are built on the
traditional, ancestral, and unceded territory of the Syilx Okanagan people.
K'ni\textipa{P}atn k'l$\left._\mathrm{\smile}\right.$stk'masqt is situated on
land leased from the Imperial Metals Corporation. We are grateful to the staff
of the Dominion Radio Astrophysical Observatory, which is operated by the
National Research Council of Canada. CHIME operations are funded by a grant
from the NSERC Alliance Program and by support from McGill University,
University of British Columbia, and University of Toronto. CHIME/FRB Outriggers
are funded by a grant from the Gordon \& Betty Moore Foundation. We are grateful
to Robert Kirshner for early support and encouragement of the CHIME/FRB
Outriggers Project, and to Dusan Pejakovic of the Moore Foundation for
continued support. CHIME was funded by a grant from the Canada Foundation for
Innovation (CFI) 2012 Leading Edge Fund (Project 31170) and by contributions
from the provinces of British Columbia, Québec and Ontario. The CHIME/FRB
Project was funded by a grant from the CFI 2015 Innovation Fund (Project 33213)
and by contributions from the provinces of British Columbia and Québec, and by
the Dunlap Institute for Astronomy and Astrophysics at the University of
Toronto. Additional support was provided by the Canadian Institute for Advanced
Research (CIFAR), the Trottier Space Institute at McGill University, and the
University of British Columbia. The CHIME/FRB baseband recording system is
funded in part by a CFI John R. Evans Leaders Fund award to IHS. 

J.M.P. acknowledges the support of an NSERC Discovery Grant (RGPIN-2023-05373).
K.W.M. holds the Adam J. Burgasser Chair in Astrophysics and is supported by an NSF Grant (2008031).
K.M.B. is supported by NSF Grant 2018490. 
M.D. is supported by a CRC Chair, NSERC Discovery Grant, and CIFAR. 
M.L. acknowledges the support of the Natural Sciences and Engineering Research Council of Canada (NSERC-CGSD)
C. L. is supported by NASA through the NASA Hubble Fellowship grant HST-HF2-51536.001-A awarded by the Space Telescope Science Institute, which is operated by the Association of Universities for Research in Astronomy, Inc., under NASA contract NAS5-26555.  
K.N. acknowledges support by NASA through the NASA Hubble Fellowship grant \# HST-HF2-51582.001-A awarded by the Space Telescope Science Institute, which is operated by the Association of Universities for Research in Astronomy, Incorporated, under NASA contract NAS5-26555.
A.B.P. acknowledges support by NASA through the NASA Hubble Fellowship grant HST-HF2-51584.001-A awarded by the Space Telescope Science Institute, which is operated by the Association of Universities for Research in Astronomy, Inc., under NASA contract NAS5-26555. A.B.P. also acknowledges prior support from a Banting Fellowship, a McGill Space Institute~(MSI) Fellowship, and a Fonds de Recherche du Quebec -- Nature et Technologies~(FRQNT) Postdoctoral Fellowship.
\appendix

\section{Toy model}
\label{sec:toy_model}
The following simple yet illustrative toy model for the unwanted interference helps us build some intuition about how the performance of the \jmp{KL beamformer} compares to a conventional beamformer. It also allows us to investigate the effect of common systematics, including calibration and pointing errors. 

We will first study the unpolarized case for a single polarization such that the signal covariance model $\matr{S}$ is given by Equation~\ref{eq:S_singlepol1}. We model the contamination covariance $\matr{F}$ as a separate point source plus uncorrelated receiver noise. The total data covariance of our toy model is then
\begin{equation}
\label{eq:tm_cov}
    \matr{D} = \langle \vect{d} \vect{d}^\dagger \rangle = \underbrace{N\sigma_s^2\vect{\hat{v}}_{s}\vect{\hat{v}}_{s}^\dagger}_{\matr{S}} + \underbrace{N\sigma_r^2\vect{\hat{v}}_{r}\vect{\hat{v}}_{r}^\dagger   + \matr{I}}_{\matr{F}}, \hspace{0.2in} v_{s,i}=\frac{e^{i2\pi \vect{u}_{i}^T\vect{\hat{n}}_s}}{\sqrt{N}}, \hspace{0.2in} v_{r,i}=\frac{e^{i2\pi \vect{u}_{i}^T\vect{\hat{n}}_{r}}}{\sqrt{N}}.
\end{equation}
The signal of interest represented as a point source in the direction of $\vect{\hat{n}}_s$, the second term is an unwanted signal represented as a point source in the direction $\vect{\hat{n}}_{r}$, and the last term represents the noise covariance. We have also assumed the antennas are identical and the receiver noise is uncorrelated with noise temperature $\sigma_n^2=1$. The covariance $\matr{D}$ is normalized such that the autocorrelation (total power) of each antenna is $D_{ii} = \sigma_s^2+\sigma_r^2 + 1$, i.e., $\sigma_s^2$ and $\sigma_r^2$ represent the per-antenna signal and contaminating source (RFI or astrophysical) power normalized by the system noise power $\sigma_n^2$. Since antennas are identical, information about the directional gains (beam) is included in $\sigma_s^2$ and $\sigma_r^2$.  

\subsection{$S/N$ measurements}
\label{ap:snr}
We are interested in comparing the $S/N$ estimates obtained with both conventional and \refereejmp{KL} beamformers. For a beamforming vector $\vect{b}$, the $S/N$ is defined by equation~\ref{eq:snr_bf}. The conventional beamformer is $\vect{b}=\vect{\hat{v}}_{s}$ (Section~\ref{sec:conv_bf}), so

\begin{equation}
\label{eq:snr_bf_toy}
    (S/N)_{BF} = \frac{\vect{\hat{v}}_{s}^\dagger \matr{S} \vect{\hat{v}}_{s}}{\vect{\hat{v}}_{s}^\dagger \matr{F} \vect{\hat{v}}_{s}} = \frac{N\sigma_s^2}{N\sigma_r^2|\vect{\hat{v}}_{s}^\dagger\vect{\hat{v}}_{r}|^2+1}.
\end{equation}

To calculate the $S/N$ of the \jmp{KL} beamformer (given by Equation~\ref{eq:klt_bf_snr}) for this toy model, we first calculate the explicit form of $\matr{F}^{-1}$. Note that $\vect{\hat{v}}_{1}=\vect{\hat{v}}_{r}$ is an eigenvector of $\matr{F}$ with eigenvalue $\lambda_1 = N\sigma_r^2+1$. The remaining eigenvectors $\left\{ \vect{\hat{v}}_{i} \right\}_{i=2,...,N}$ have the same eigenvalue $\lambda_i=1$ and form an orthonormal basis of the subspace orthogonal to $\vect{\hat{v}}_{r}$. $\matr{F}$ is hermitian and positive definite. We can write
\begin{align}
    \matr{F} &= \sum_i  \lambda_i \vect{\hat{v}}_{i} \vect{\hat{v}}_{i}^\dagger = (N\sigma_r^2+1)\vect{\hat{v}}_{r}\vect{\hat{v}}_{r}^\dagger   + (\matr{I}-\vect{\hat{v}}_{r}\vect{\hat{v}}_{r}^\dagger) \\
    \matr{F}^{-1} &= \sum_i  \frac{1}{\lambda_i} \vect{\hat{v}}_{i} \vect{\hat{v}}_{i}^\dagger = \frac{1}{N\sigma_r^2+1}\vect{\hat{v}}_{r}\vect{\hat{v}}_{r}^\dagger   + (\matr{I}-\vect{\hat{v}}_{r}\vect{\hat{v}}_{r}^\dagger)= \matr{I}-\frac{N\sigma_r^2}{N\sigma_r^2+1}\vect{\hat{v}}_{r}\vect{\hat{v}}_{r}^\dagger \\
\label{eq:snr_klt}
    (S/N)_{KLT} &= N\sigma_s^2\vect{\hat{v}}_{s}^\dagger \matr{F}^{-1} \vect{\hat{v}}_{s} = N\sigma_s^2\left(1-\frac{N\sigma_r^2|\vect{\hat{v}}_{s}^\dagger\vect{\hat{v}}_{r}|^2}{N\sigma_r^2+1}\right).
\end{align}
This provides
\begin{equation}
\label{eq:snr_ratio}
    \frac{(S/N)_{KLT}}{(S/N)_{BF}} = 1+ N\sigma_r^2|\vect{\hat{v}}_{s}^\dagger\vect{\hat{v}}_{r}|^2 \left (1- \frac{1+ N\sigma_r^2|\vect{\hat{v}}_{s}^\dagger\vect{\hat{v}}_{r}|^2}{1+ N\sigma_r^2} \right ).
\end{equation}
Note that since $0\leq |\vect{\hat{v}}_{s}^\dagger\vect{\hat{v}}_{r}|^2\leq 1$, it follows that $(S/N)_{KLT}/(S/N)_{BF}\geq 1$. This is expected since the \refereejmp{KL} beamformer, by definition, is constructed to maximize $S/N$ (Section~\ref{sec:klt_bf}). 

\begin{figure}[h]
    \centering
    \includegraphics[width=\linewidth]{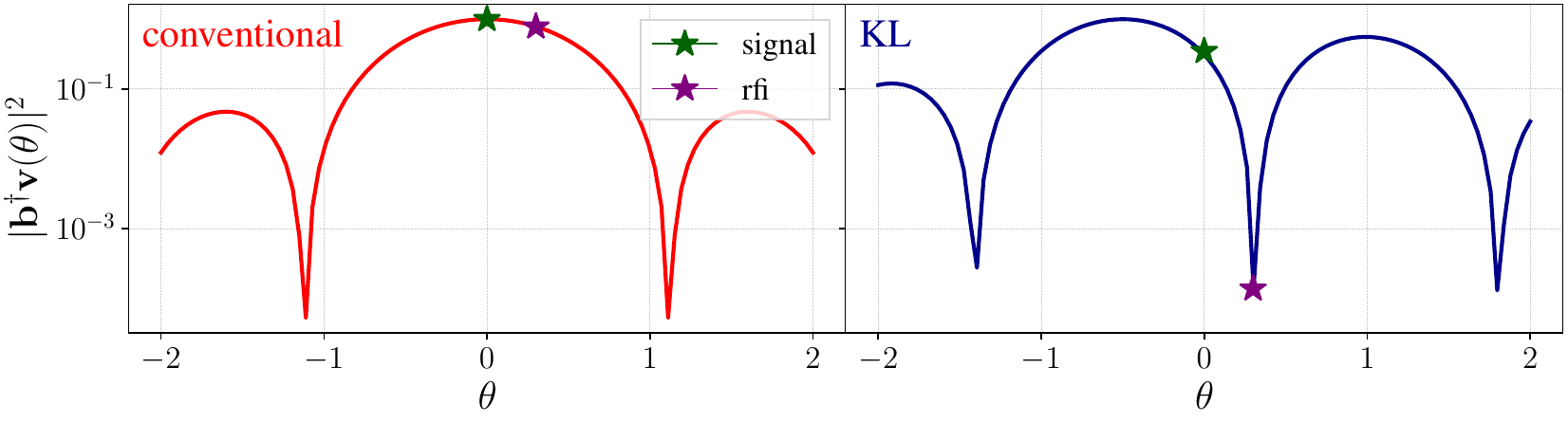}
    \caption{\fix{Normalized beam response to a source in the sky $\vect{v}(\theta)$, or equivalently $|b^\dagger v(\theta)|^2$ where the beam weights $b$ are calculated for $\theta_s=0$ and $\theta_r=0.3^\circ$. Left panel: the normalized beam response of the conventional beamformer, where the locations of the signal and RFI are indicated by the green and purple stars, respectively. Note that although the beam response is maximized towards the direction of the signal, the signal-to-noise ratio is determined by the relative response of the beam at the location of the signal to the response of the beam at the location of the RFI. Right panel: the normalized beam response of the KL beamformer. While the beam response here is no longer maximized towards the signal, the beamformer by design is constructed to maximize the signal-to-noise ratio, which is clearly larger than the signal-to-noise ratio for the conventional beamformer by several orders of magnitude.}}\label{fig:beam_response}
\end{figure}

\begin{figure}[h]
    \centering
    \includegraphics[width=\linewidth]{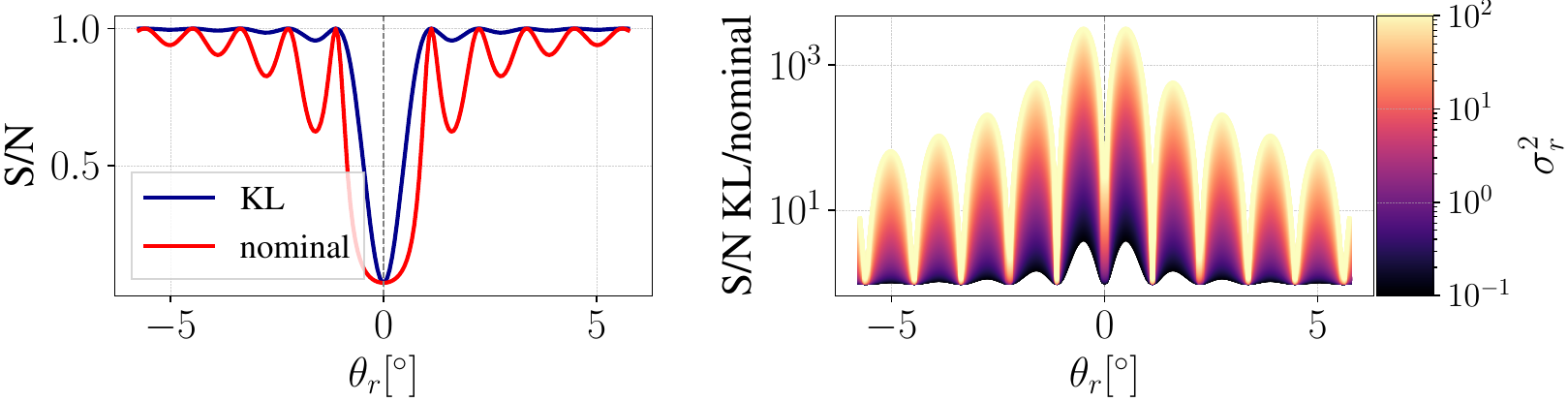}
    \caption{$S/N$s as function of RFI location ($\theta_r$) for a 128-element regularly spaced 1D array with $x_\lambda=0.4$ (CHIME Outrigger at 400 MHz) \fix{and $\sigma_s^2=0.1$. We parameterize the location of the point-like RFI (target) with $\theta_r$ ($\theta_s$) where $\vect{u}_{j}^T\vect{\hat{n}}_{r}=jx_\lambda\sin\theta_r$ ($\vect{u}_{j}^T\vect{\hat{n}}_s=jx_\lambda\sin\theta_s$). We place the signal at zenith ($\theta_s=0^\circ$), which is indicated with the dashed line.} Left panel: comparison of the $S/N$ in the conventional vs KL beamformer for the case where \fix{$\sigma_r^2=1$}. While the performance of both beamformers drop as the RFI mode overlaps the signal mode, the KL beamformer is notably more robust. Right panel: the ratios of the signal-to-noises as a function of $\sigma_r^2$ and $\theta_r$ with a dashed line drawn at a ratio of one. The signal-to-noise gain seen by the KL filter increases sharply with RFI power. \secondfix{We also see that if the RFI is placed in one of the nulls of the conventional synthesized beam, the RFI contribution to the total beamformed power is very small and the KL filter sees minimal improvement: this corresponds to the local maxima in the left panel. The KL filter sees the most improvement when the RFI is placed in the middle of a ``sidelobe" of the conventional synthesized beam, corresponding to the local maxima in the left panel.}}
    \label{fig:snr_ratio_model}
\end{figure}

\fix{To visualize Equation \ref{eq:snr_ratio}, we now consider the case for an Outriggers-like, regularly spaced 1D array with parallel baseline vectors for all pairs of antenna elements. We parameterize the location of our source (RFI) with the parameter $\theta_s$ ($\theta_r$) where $v_{s,j} = e^{i2\pi j x_\lambda \sin\theta_s}/\sqrt{N}$ and $x_\lambda$ is the array spacing in wavelengths (we adopt a value of 0.4 for CHIME). We also place the FRB at zenith $(\theta_s = 0^\circ)$.
To build intuition, we first plot the normalized beam response for the conventional and KL beamformers in Figure \ref{fig:beam_response}. For the conventional beamformer, the beam is determined completely by $\theta_s$, which we have set here to $0^\circ$. For the KL beamformer, the beam is also determined by the location of the RFI, which we have chosen here to be placed at $\theta_r=0.3^\circ$. 

Next, we vary the placement of the RFI as well as its brightness. The resulting signal-to-noise distribution for the KL relative to the conventional beamformer in the case of this specific toy model is illustrated in Figure \ref{fig:snr_ratio_model}. In accordance with \ref{eq:snr_ratio}, we see that the KL filter performs better relative to the nominal beamformer the brighter (larger $\sigma_r^2$) the RFI.} \secondfix{The performance improvement also depends on how well the interference and astrophysical signal occupy distinct spatial modes: when the RFI lies in a spatial mode nearly orthogonal to the signal response, the KL filter can strongly suppress the interference while preserving the signal.}

\subsection{Calibration Errors}
\label{subsec:phase_errors}
The statement about the \jmp{KL} beamformer maximizing $S/N$ assumes that both \jmp{$\matr{S}$} and \jmp{$\matr{F}$} are known. In practice, we typically have a good initial guess but do not know these terms perfectly for reasons such as calibration errors, baseline errors, and pointing errors. Here, we are interested in investigating the robustness of the \jmp{KL beamformer} against these systematics and compare its performance to that of the conventional CHIME/FRB beamformer. 
Of particular interest are errors in the calibration of the per-antenna amplitude and phase response. This per-antenna response is calibrated by observing bright steady sources, and is typically known at the percent level at CHIME ($\sim 0.01$ rad for phases) within the clean parts of the band \citep{chime_overview}. However, the CHIME \jmp{real-time processing} pipeline does not directly generate calibration gains for frequency channels that are permanently contaminated by \jmp{bright} RFI \jmp{since the application of the default calibration algorithm results in biased gains. Instead,} we must obtain solutions offline by interpolating gains from clean neighboring channels. This interpolation operation can result in gain errors well above $\sim$10\%, especially across wide RFI bands where the interpolation is large (several 10s of MHz).

We can model phase errors in the calibration solution as an antenna-dependent multiplicative phase term $d_{i} \rightarrow e^{i\delta \phi_i} d_{i}$ that is constant during the observation of the transient source. When applying this error to the toy model described in Section \ref{sec:toy_model}, the model for our data covariance becomes
\begin{equation}
\label{eq:tm_cov_phase_err}
    \tilde{\matr{D}} = \matr{\Phi}\matr{D}\matr{\Phi}^\dagger=\matr{\Phi}(\matr{S}+\matr{F})\matr{\Phi}^\dagger
\end{equation}
where $\matr{\Phi}=\text{diag}(e^{i\delta \phi_0}, e^{i\delta \phi_1},..., e^{i\delta \phi_{N-1}})$. We do not know these residual terms, but typically have an idea about their scale and statistics. In the simplest case we assume $\langle \delta\phi_i \rangle_\phi = 0$ and  $\langle \delta\phi_i \delta\phi_j \rangle_\phi = \sigma_\phi ^2 \delta_{ij}$. Here, $\langle \cdot \rangle_\phi$ denotes average over phase error realizations. \refereejmp{In the analysis that follows, this averaging operation is applied separately to the beamformed signal and noise power. This extends the definition in Equation~\ref{eq:snr_bf}, which is a ratio of two time-averaged powers, to an enlarged ensemble including time and phase errors.} We will assume that, although the calibration is imperfect, the position of the source of interest is known. This means that the conventional beamformer is still $\vect{b} = \vect{\hat{v}}_{s}$ and

 

\begin{equation}
    \tilde{\matr{F}} = \matr{\Phi}\matr{F}\matr{\Phi}^\dagger=N\sigma_r^2\matr{\Phi}\vect{\hat{v}}_{r}\vect{\hat{v}}_{r}^\dagger \matr{\Phi}^\dagger   + \matr{I}  \hspace{0.2in} \rightarrow \hspace{0.2in} \langle \vect{\hat{v}}_{s}^\dagger \tilde{\matr{F}} \vect{\hat{v}}_{s} \rangle_\phi = N\sigma_r^2 \langle |\vect{\hat{v}}_{s}^\dagger \matr{\Phi} \vect{\hat{v}}_{r}|^2\rangle_\phi +1
\end{equation}
where

\begin{equation}
    \langle |\vect{\hat{v}}_{s}^\dagger \matr{\Phi} \vect{\hat{v}}_{r}|^2\rangle_\phi = \sum_{ij} \langle e^{i\delta \phi_i} e^{-i\delta \phi_j} \rangle_\phi v_{s,i}^*v_{r,i} v_{s,j}v_{r,j}^*
\end{equation}
A closed-form expression can be found in the case where the phase errors are Gaussian \jmp{(a reasonable approximation if the errors are smaller than about a radian)}
\begin{align}
    \langle e^{i\delta \phi_i} e^{-i\delta \phi_j} \rangle_\phi &= \begin{cases}
     1  & \text{if } i=j \\
     \langle e^{i\delta \phi_i} \rangle_\phi \langle e^{-i\delta \phi_j} \rangle_\phi = e^{-\sigma_\phi^2} & \text{if } i\neq j 
\end{cases}\\
    &= e^{-\sigma_\phi^2} + (1-e^{-\sigma_\phi^2})\delta_{ij}.
\end{align}
Thus we have
\begin{align}
    \langle |\vect{\hat{v}}_{s}^\dagger \matr{\Phi} \vect{\hat{v}}_{r}|^2\rangle_\phi &= e^{-\sigma_\phi^2} |\vect{\hat{v}}_{s}^\dagger  \vect{\hat{v}}_{r}|^2 + \frac{1-e^{-\sigma_\phi^2}}{N}\\
    \langle \vect{\hat{v}}_{s}^\dagger \tilde{\matr{F}} \vect{\hat{v}}_{s} \rangle_\phi &= N\sigma_r^2 \left (e^{-\sigma_\phi^2} |\vect{\hat{v}}_{s}^\dagger  \vect{\hat{v}}_{r}|^2 + \frac{1-e^{-\sigma_\phi^2}}{N} \right ) +1\\
    \langle \vect{\hat{v}}_{s}^\dagger \tilde{\matr{S}} \vect{\hat{v}}_{s} \rangle_\phi &= N\sigma_s^2 \langle |\vect{\hat{v}}_{s}^\dagger \matr{\Phi} \vect{\hat{v}}_{s}|^2\rangle_\phi = N\sigma_s^2 \left (e^{-\sigma_\phi^2} + \frac{1-e^{-\sigma_\phi^2}}{N} \right ).
\end{align}
This provides us with the average $(S/N)_{BF}$: 
\begin{equation}
\label{eqn:conventional_snr_phase}
    (S/N)_{BF,\phi} = N\sigma_s^2 \cdot \frac{\displaystyle e^{-\sigma_\phi^2} + \frac{1-e^{-\sigma_\phi^2}}{N}}{\displaystyle N\sigma_r^2 \left (e^{-\sigma_\phi^2} |\vect{\hat{v}}_{s}^\dagger  \vect{\hat{v}}_{r}|^2 + \frac{1-e^{-\sigma_\phi^2}}{N} \right ) +1}.
\end{equation}

In the case of the \jmp{KL} beamformer, we estimate \jmp{the} covariance of the contaminant source and noise from the (now miscalibrated) off-pulse data. \jmp{The} beamformer \jmp{will have} the form $\vect{b} \propto \tilde{\matr{F}}^{-1}\vect{\hat{v}}_{s}$, where $\tilde{\matr{F}}^{-1}=\matr{\Phi}\matr{F}^{-1}\matr{\Phi}^\dagger$ \jmp{since $\matr{\Phi}$ is diagonal consisting of phase terms only}. Thus
\begin{align}
    \langle \vect{b}^\dagger \tilde{\matr{F}} \vect{b} \rangle_\phi &= \langle \vect{\hat{v}}_{s}^\dagger \tilde{\matr{F}}^{-1} \vect{\hat{v}}_{s} \rangle_\phi \nonumber \\
    &= 1- \frac{N\sigma_r^2}{N\sigma_r^2+1} \langle |\vect{\hat{v}}_{s}^\dagger \matr{\Phi}  \vect{\hat{v}}_{r}|^2 \rangle_\phi \\
     \langle \vect{b}^\dagger \tilde{\matr{S}} \vect{b} \rangle_\phi &= \langle \vect{\hat{v}}_{s}^\dagger \matr{\Phi}\matr{F}^{-1}\matr{S} \matr{F}^{-1}\matr{\Phi}^\dagger \vect{\hat{v}}_{s} \rangle_\phi \nonumber\\
     &= N\sigma_s^2\left \langle  |\vect{\hat{v}}_{s}^\dagger \matr{\Phi}  \vect{\hat{v}}_{s}|^2 - \frac{N\sigma_r^2}{N\sigma_r^2+1} (\vect{\hat{v}}_{s}^\dagger \matr{\Phi}  \vect{\hat{v}}_{s} \vect{\hat{v}}_{s}^\dagger  \vect{\hat{v}_r \vect{\hat{v}}_{r}^\dagger \matr{\Phi}^\dagger  \vect{\hat{v}}_{s}} + \vect{\hat{v}}_{s}^\dagger \matr{\Phi}  \vect{\hat{v}}_{r} \vect{\hat{v}}_{r}^\dagger  \vect{\hat{v}_s \vect{\hat{v}}_{s}^\dagger \matr{\Phi}^\dagger  \vect{\hat{v}}_{s}}) + \left(  \frac{N\sigma_r^2}{N\sigma_r^2+1} \right)^2 |\vect{\hat{v}}_{s}^\dagger   \vect{\hat{v}}_{r}|^2|\vect{\hat{v}}_{s}^\dagger \matr{\Phi}  \vect{\hat{v}}_{r}|^2 \right \rangle_\phi
\end{align}
which provides
\begin{equation}
\label{eq:snr_klt_phase_error2}
    (S/N)_{KLT,\phi} = N\sigma_s^2 \cdot \frac{\displaystyle \left (e^{-\sigma_\phi^2} + \frac{1-e^{-\sigma_\phi^2}}{N}\right) \left( 1-\frac{N\sigma_r^2|\vect{\hat{v}}_{s}^\dagger\vect{\hat{v}}_{r}|^2}{N\sigma_r^2+1} \right)^2 + \left( \frac{N\sigma_r^2}{N\sigma_r^2+1} \right)^2 \frac{\left(1-e^{-\sigma_\phi^2}\right)}{N}|\vect{\hat{v}}_{s}^\dagger\vect{\hat{v}}_{r}|^2 \left(1-|\vect{\hat{v}}_{s}^\dagger\vect{\hat{v}}_{r}|^2 \right)}
    {\displaystyle 1- \left( \frac{N\sigma_r^2}{N\sigma_r^2+1} \right) \left (e^{-\sigma_\phi^2} |\vect{\hat{v}}_{s}^\dagger  \vect{\hat{v}}_{r}|^2 + \frac{1-e^{-\sigma_\phi^2}}{N} \right )}.
\end{equation}

\begin{figure}[h]
    \centering
    \includegraphics[width=\linewidth]{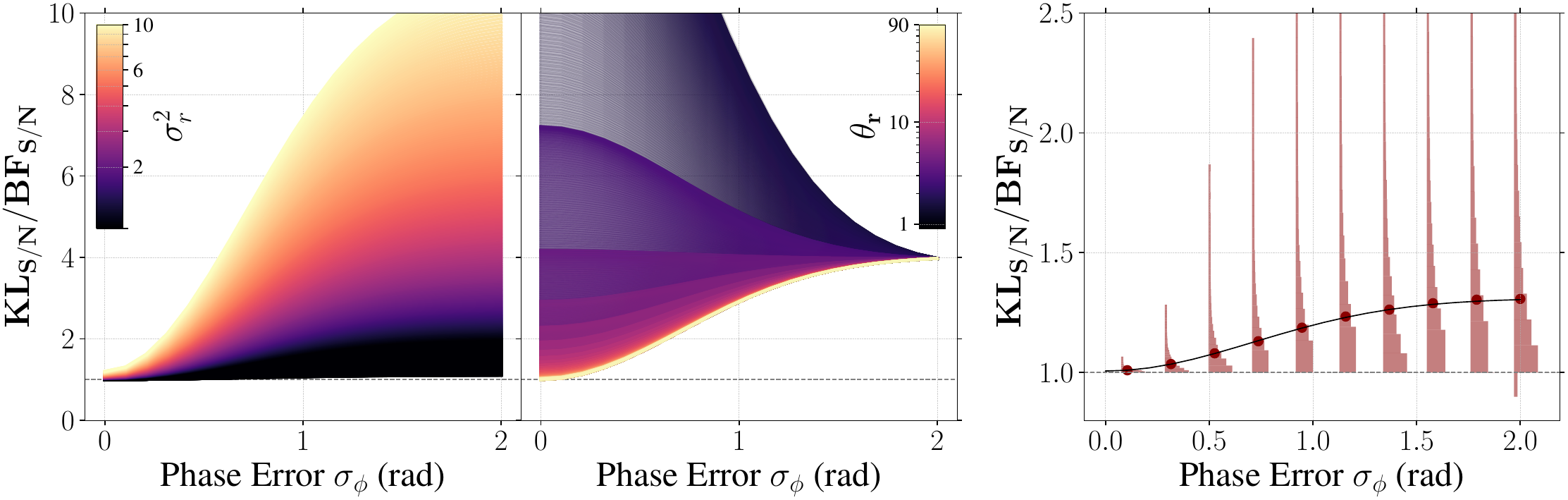}
    \caption{Ratio of $S/N$s according to Equations \ref{eqn:conventional_snr_phase} and \ref{eq:snr_klt_phase_error2}  as a function of the error in phase $\sigma_\phi$ for a 128-element regularly spaced 1D array with \jmp{$x_\lambda=0.4$} (Outrigger at 400 MHz). Left panel: ratio of \jmp{the $S/N$ of the KL beamformer} to the $S/N$ of the conventional beamformer for different values of $\sigma_r^2$ with the RFI location fixed at  $\theta_r=30^\circ$. Middle panel: $S/N$ ratios for different values of $\theta_r$ with $\sigma_r^2$ fixed at $\sigma_r^2=3$. Right panel: the average $S/N$ ratio \finalfix{(red dots)} over specific realizations of $\delta\phi$ randomly drawn in simulations with \finalfix{the distribution of $S/N$ ratios for each $\sigma_\phi$ shown as a histogram} compared to the model curve \finalfix{(solid line)} in the case of $\sigma_r^2=3,\theta_r=30^\circ$.  We see good agreement \finalfix{between the average of the distribution and the model curve}. \finalfix{Note, however, that while the distribution itself is highly asymmetric, the signal-to-noise of the KL filter does not drop below the signal-to-noise of the traditional beamformer except in the limit that $\sigma_\phi$ is sufficiently large (exceeds $\sim$2 rad in this particular example).  }}\label{fig:snr_ratio_model_phase_error} 
\end{figure}
As shown in Figure \ref{fig:snr_ratio_model_phase_error}, we find that the simulations of this toy model are consistent with the prediction from Equation~\ref{eq:snr_klt_phase_error2}. \finalfix{Over all realizations of $\delta \phi$, the \refereejmp{KL} filter performs better than the conventional beamformer in the presence of phase calibration errors, provided they are sufficiently small.}








In the case of amplitude errors, if we parametrize the gain magnitudes as $e^{g_i}$, with the $g_i$ as real valued, then the model for amplitude perturbations is similar to that of section~\ref{subsec:phase_errors}, with $\matr{\Phi}=\text{diag}(e^{\delta g_0}, e^{\delta g_1},..., e^{\delta g_{N-1}})$. In addition, we have:

\begin{equation}
    \tilde{\matr{F}}^{-1}=\matr{\Phi}^{-1}\matr{F}^{-1}\matr{\Phi}^{-1}, \hspace{0.2in} \langle e^{\delta g_i} e^{\delta g_j} \rangle_g = e^{\sigma_g^2} + (e^{2\sigma_g^2}-e^{\sigma_g^2})\delta_{ij} = e^{2\sigma_g^2} \left[e^{-\sigma_g^2} + (1-e^{-\sigma_g^2})\delta_{ij}\right]
\end{equation}

This results in expressions for $(S/N)_{BF,g}$ and \textbf{$(S/N)_{KLT,g}$} that are identical to those of phase errors, with the replacement $\sigma_\phi \rightarrow \sigma_g$.

\subsection{Pointing errors}
\label{subsec:Pointingerrors}

We now consider the case that calibration is correct but the source direction $\theta_s$ is incorrectly modeled as $\tilde{\theta}_s=\theta_s+\delta\theta$ where $\delta\theta$ follows the Gaussian distribution $\mathcal{N}=(0,\sigma_\theta)$. This means that beamformer $\vect{b}$ is perturbed due to the imperfect knowledge of $\vect{\hat{v}}_{s}$. \fix{As before, we consider a single CHIME cylinder and take $v_{s,j} = e^{i2\pi j x_\lambda \sin\theta_s}/\sqrt{N}$ for simplicity.} For a small perturbation $\delta\theta\ll1$ to the modeled source direction, a perturbed conventional beamformer takes the form $\vect{b}=\matr{\Phi}\vect{\hat{v}}_{s}$, where 
\begin{equation}
\Phi_{kl} = e^{i2\pi k x_\lambda \cos\theta_s \delta\theta}\delta_{kl} = e^{i k \delta\gamma
}\delta_{kl}
\end{equation}
\jmp{where $\delta \gamma = 2\pi x_\lambda \cos\theta_s\delta \theta$.} This provides
\begin{align}
\langle \vect{b}^\dagger \matr{F} \vect{b} \rangle_\gamma
 &= N\sigma_r^2 \langle |\vect{\hat{v}}_{s}^\dagger \matr{\Phi}\fix{^\dagger} \vect{\hat{v}}_{r}|^2\rangle_\gamma
 +1\\
\langle |\vect{\hat{v}}_{s}^\dagger \matr{\Phi}\fix{^\dagger} \vect{\hat{v}}_{r}|^2\rangle_\gamma
 &= \sum_{kl} \langle e^{\fix{-}i (k-l) \delta\gamma
}\rangle_\gamma
 v_{s,k}^*v_{r,k} v_{s,l}v_{r,l}^*\\
&= \frac{1}{N^2} \sum_{kl} e^{- (k-l)^2 \sigma_\gamma
^2/2} e^{-i2\pi k x_\lambda(\sin\theta_s-\sin\theta_r)} e^{i2\pi l x_\lambda(\sin\theta_s-\sin\theta_r)},
\end{align}
where $\sigma_\gamma
=2\pi x_\lambda \cos\theta_s\sigma_\theta$. The expression above looks like the truncated version of a 2D discrete-time Fourier Transform (DTFT).
If we define the function $\mathcal{R}(y,z)$ as
\begin{equation}
\label{eq:R_equation}
\mathcal{R}_\gamma
(y,z)=\frac{1}{N^2}DTFT\left\{ e^{- (k-l)^2 \sigma_\gamma
^2/2} w(k,l)\right\}(y, z), \hspace{.2in} w(k, l) = \begin{cases}
     1  & \text{if } 0\leq k,l \leq N-1 \\
     0 & \text{else}
\end{cases}
\end{equation}
Then 
\begin{align}
\langle \vect{b}^\dagger \matr{F} \vect{b} \rangle_\gamma
 &= N\sigma_r^2 \mathcal{R}_\gamma
(u, -u) +1, \hspace{.2in} u=x_\lambda(\sin\theta_s-\sin\theta_r) \\    \langle \vect{b}^\dagger \fix{\matr{S}} \vect{b} \rangle_\refereejmp{\gamma} &= N\sigma_s^2 \langle |\vect{\hat{v}}_{s}^\dagger \matr{\Phi}\fix{^\dagger} \vect{\hat{v}}_{s}|^2\rangle_\gamma
 = N\sigma_s^2 \mathcal{R}_\gamma
(0, 0),
\end{align}
which provides 
\begin{equation}
\label{eqn:BF_pointing_ratio}
    (S/N)_{BF,\gamma
} = N\sigma_s^2 \cdot \frac{\mathcal{R}_\gamma
(0, 0)}{N\sigma_r^2 \mathcal{R}_\gamma
(u, -u) +1}.
\end{equation}

\begin{figure}[h]
    \centering
    \includegraphics[width=\linewidth]{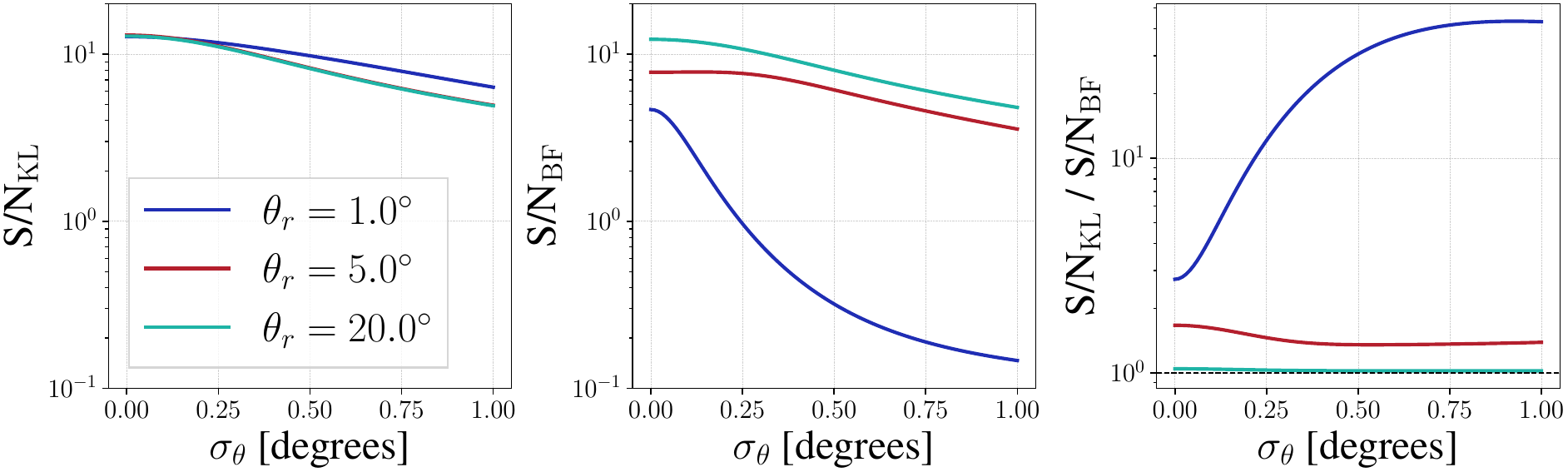}
\caption{S/Ns according to Equations \ref{eqn:BF_pointing_ratio} and \ref{eqn:KL_pointing_ratio}  as a function of the pointing error $\sigma_\theta$ \fix{with $\sigma_s=0.1,\sigma_r=1.0,$ and $\theta_s=0$}  In both beamformers, the S/N degrades as the pointing error increases, and the ratio of the S/Ns also decreases except if the RFI is within $\sim$ one synthesized beam of the target (blue curve). \label{fig:pointing_error}} 
\end{figure}

In the case of the \refereejmp{KL} filter, the beamformer has the form $\vect{b} = \fix{\matr{F}}^{-1}\matr{\Phi}\vect{\hat{v}}_{s}$. A similar calculation gives

\begin{equation}
\label{eqn:KL_pointing_ratio}
    (S/N)_{KLT,\gamma
} = N\sigma_s^2 \cdot \frac{\displaystyle \mathcal{R}_\gamma
(0, 0)-  \frac{2N\sigma_r^2}{N\sigma_r^2+1} \Re\left\{\vect{\hat{v}}_{s}^\dagger\vect{\hat{v}}_{r} \mathcal{R}_\gamma
(0, -u) \right\}+\left( \frac{N\sigma_r^2}{N\sigma_r^2+1} \right)^2|\vect{\hat{v}}_{s}^\dagger  \vect{\hat{v}}_{r}|^2 \mathcal{R}_\gamma
(u, -u)}{\displaystyle  1-\jmp{\frac{N\sigma_r^2}{N\sigma_r^2+1}} \mathcal{R}_\gamma
(u, -u)}
\end{equation}
\fix{where $ \Re \{x\}$ denotes the real part of the number $x$}. Figure \ref{fig:pointing_error} compares the signal-to-noise performance of the conventional and KL beamformer for different pointing errors according to Equations \ref{eqn:BF_pointing_ratio} and \ref{eqn:KL_pointing_ratio}. \fix{The performance of both the conventional and KL filter degrade with larger pointing errors, and depending on the location of the RFI relative to the target of interest, can reduce the relative signal-to-noise gain seen by the KL filter over the conventional beamformer.}

\section{Extensions}
\label{ap:extensions}
In this section, we briefly discuss variations of the KL filter following the more general form given by Equation \ref{eq:klt_eval_problem}.
\subsection{Polarimetric \jmp{KLT} Filter}
\label{sec:polarization}
Because RFI is often highly polarized, including polarimetry in the \jmp{KL} filter should result in a much cleaner separation of RFI modes from the signal. Here we assume that the polarization of the source is known and discuss the \jmp{implementation of a multi-mode \refereejmp{KL} filter for polarization, its performance, and the constraints it provides on the optimal total $S/N$ discussed in Section~\ref{sec:klt}.}


To start, let us look at the response of one of the dual-polarization antennas of our array to the point source signal in Equation~\ref{eq:vis_ps}. If the $i$-th and $j$-th elements form a dual-polarization antenna pair, then the fringing term in  Equation~\ref{eq:vis_ps} is unity. Let $i$ be mostly sensitive to the $X$ polarization and $j$ to the $Y$ polarization of emission. If the polarization leakage is negligible, then

\begin{align}
\label{eq:dual_pol_coherence_mat_beam}
   &\left\langle \begin{bmatrix}
d_i \\
d_j 
\end{bmatrix} \begin{bmatrix}
d_i^* & d_j^*
\end{bmatrix}\right \rangle  = \frac{1}{2} \begin{bmatrix}
|A_i^X(\vect{\hat{n}}_s)|^2(T_s+Q_s) & A_i^X(\vect{\hat{n}}_s)A_j^{Y*}(\vect{\hat{n}}_s)(U_s-iV_s)\\
A_j^Y(\vect{\hat{n}}_s)A_i^{X*}(\vect{\hat{n}}_s)(U_s+iV_s) & |A_j^Y(\vect{\hat{n}}_s)|^2(T_s-Q_s)
\end{bmatrix}.
\end{align}

If in addition, the antenna beams are calibrated out, we have

\begin{equation}
\label{eq:dual_pol_coherence_mat}
   \left\langle \begin{bmatrix}
d_i \\
d_j 
\end{bmatrix} \begin{bmatrix}
d_i^* & d_j^*
\end{bmatrix}\right \rangle =\matr{C}(\vect{\hat{n}}_s) = \frac{1}{2} \begin{bmatrix}
T_s+Q_s & U_s-iV_s\\
U_s+iV_s & T_s-Q_s
\end{bmatrix}
\end{equation}
where $(T_s,~Q_s,~U_s,~V_s)$ are the Stokes parameters of the source. This matrix by construction is hermitian and positive semi-definite, and has eigenvalues
\begin{equation}
\label{eq:Ts_evalues}
   \lambda_{\pm} = \frac{1}{2}(T_s\pm T_{\jmp{ps}}), \hspace{0.2in} T_{\jmp{ps}}^2 = Q_s^2+U_s^2+V_s^2 \leq T_s^2
\end{equation}
where $T_{\jmp{ps}}$ is the source polarized intensity. The respective eigenvectors are

\begin{equation}
\label{eq:Ts_evalues}
   \vect{\hat{v}_+} = \begin{bmatrix}
e^{-i\chi_s}\cos(\psi_s/2) \\
\sin(\psi_s/2) 
\end{bmatrix}, \hspace{0.2in} \vect{\hat{v}_-} = \begin{bmatrix}
-e^{-i\chi_s}\sin(\psi_s/2) \\
\cos(\psi_s/2) 
\end{bmatrix}
\end{equation}
where $\tan \chi_s=\refereejmp{V_s/U_s}$ and $\cos \psi_s = Q_s/\fix{T}_{ps}$.

Returning to the response of the full array, if we choose our antenna index mapping $i$ such that the first subset of $N$ antenna elements is $X$ polarized and the second subset is $Y$ polarized, then the point source visibility matrix can be written in block matrix form as

\begin{align}
\label{eq:S_pol}
   \matr{S} &= \begin{bmatrix}
\matr{S}^{XX} & \matr{S}^{XY}\\
\matr{S}^{YX} & \matr{S}^{YY}
\end{bmatrix} \\ \label{eq:S_pol1}
&= \frac{N}{2} \begin{bmatrix}
(T_s+Q_s)\vect{\hat{v}}_{s}^X\vect{\hat{v}}_{s}^{X\dagger} & (U_s-iV_s)\vect{\hat{v}}_{s}^X\vect{\hat{v}}_{s}^{Y\dagger}\\
(U_s+iV_s)\vect{\hat{v}}_{s}^Y\vect{\hat{v}}_{s}^{X\dagger} & (T_s-Q_s)\vect{\hat{v}}_{s}^Y\vect{\hat{v}}_{s}^{Y\dagger}
\end{bmatrix}
\end{align}
where $\vect{\hat{v}}_{s}^X$ and $\vect{\hat{v}}_{s}^Y$ are defined as in Equation~\ref{eq:S_singlepol1} for the $X$ and $Y$ polarized antenna elements respectively. Note that if we choose our indexing such that the $i$-th and $N+i$-th array elements correspond to the two elements of the same dual-polarization antenna we have $\vect{\hat{v}}_{s}^X=\vect{\hat{v}}_{s}^Y$.

It is more useful for our KLT analysis to write $\matr{S}$ in terms of its eigendecomposition. In this form, the total data covariance can be written as

\begin{align}
\label{eq:D_pol}
   \matr{D} &= \frac{N}{2} (T_s+T_{ps})\vect{\hat{v}}_{s1}\vect{\hat{v}}_{s1}^\dagger + \frac{N}{2} (T_s-T_{ps})\vect{\hat{v}}_{s2}\vect{\hat{v}}_{s2}^\dagger+\matr{F}
\end{align}
where

\begin{equation}
\label{eq:Ts_evalues}
   \vect{\hat{v}}_{s1} = \begin{bmatrix}
e^{-i\chi_s}\cos(\psi_s/2)\vect{\hat{v}}_{s}^X \\
\sin(\psi_s/2) \vect{\hat{v}}_{s}^Y
\end{bmatrix}, ~ \vect{\hat{v}}_{s2} = \begin{bmatrix}
-e^{-i\chi_s}\sin(\psi_s/2)\vect{\hat{v}}_{s}^X \\
\cos(\psi_s/2) \vect{\hat{v}}_{s}^Y
\end{bmatrix}.
\end{equation}

Note that $\matr{S}$ is at most rank 2. Even with Equation~\ref{eq:dual_pol_coherence_mat_beam} when the antenna beams are not known but the polarization leakage is negligible the matrix $\matr{S}$ can be written in a similar form using the transformation shown in case 3 of Section~\ref{sec:conv_bf} and re-scaling the eigenvalues of $\matr{S}$. In the more general case where the beams are unknown and there is polarization leakage, \jmp{$\matr{S}$ will also be at most rank 2, but each block matrix in Equation~\ref{eq:S_pol} will include combinations of all the Stokes \refereejmp{parameters}}. For the purpose of demonstrating the KLT analysis we will use Equation~\ref{eq:D_pol}.

First, note that if the source is completely polarized we have $T_{ps}=T_\jmp{s}$ and $\matr{S}$ becomes rank 1 with a single non-zero eigenvalue $NT_s$ and respective eigenvector $\vect{\hat{v}}_{s1}$. This problem can be solved using the methods of Sections~\ref{sec:conv_bf} and \ref{sec:klt_bf}. In the more general case that the signal is partially polarized $\matr{S}$ will be rank 2.

It can be shown \citep{Shaw_2015} that the KLT can be implemented as a sequence of three linear transformations

\begin{equation}
\label{eq:B}
   \matr{B}=\matr{P}_{F}\matr{\Sigma}_{F}^{-1/2}\fix{\matr{P}}
\end{equation}
where $\matr{P}_{F}$ and $\matr{\Sigma}_{F}$ are the unitary matrix that diagonalizes $\matr{F}$ and its respective diagonal eigenvalue  factorization. That is

\begin{equation}
\label{eq:F}
\matr{P}_{F}^\dagger  \matr{F}\matr{P}_{F}=\matr{\Sigma}_{F}.
\end{equation}
\jmp{Note that the matrix $\matr{\Sigma}_{F}^{-1/2}$ is well defined because $\matr{F}$ is hermitian and positive definite, so all its eigenvalues are positive.}

If we apply the first two transformations to the data we obtain

\begin{equation}
\label{eq:D1}
\vect{d'} = \matr{P}_{F}^\dagger  \vect{d} \rightarrow \matr{D'}=\matr{P}_{F}^\dagger\matr{D}\matr{P}_{F}= \matr{P}_{F}^\dagger\matr{S}\matr{P}_{F}+\matr{\Sigma}_{F}
\end{equation}

\begin{align}
\label{eq:D2}
\vect{d''} = \matr{\Sigma}_{F}^{-1/2}  \vect{d'} \rightarrow \matr{D''}&=\matr{\Sigma}_{F}^{-1/2}\matr{D'}\matr{\Sigma}_{F}^{-1/2}\\\nonumber
&= \matr{\Sigma}_{F}^{-1/2} \matr{P}_{F}^\dagger\matr{S}\matr{P}_{F}\matr{\Sigma}_{F}^{-1/2} +\matr{I}\\\nonumber
&=\matr{S''} +\matr{I}
\end{align}

\jmp{The last transformation $\matr{P}$ in Equation~\ref{eq:B} is a unitary matrix that diagonalizes $\matr{S''}$, that is 

\begin{equation}
\label{eq:S2}
\matr{P}^\dagger  \matr{S''}\matr{P}=\matr{\Lambda}.
\end{equation}}

Applying  this last transformation to $\vect{d''}$  leads to Equation~\ref{eq:klt_diagonalized_cov}. Instead of doing that, let us check what $\matr{D''}$ looks like for our partially polarized example

\begin{align}
\label{eq:D2_expanded}
   \matr{D''} &= \underbrace{|\vect{\alpha}_{1}|^2\vect{\hat{\alpha}}_{1}\vect{\hat{\alpha}}_{1}^\dagger + |\vect{\alpha}_{2}|^2\vect{\hat{\alpha}}_{2}\vect{\hat{\alpha}}_{2}^\dagger}_{\matr{S''}}+\matr{I}
\end{align}
where 

\begin{align}
\label{eq:alpha1}
\vect{\alpha}_{1}&=\sqrt{\frac{N}{2}(T_s+T_{ps})}~\matr{\Sigma}_{F}^{-1/2} \matr{P}_{F}^\dagger \vect{\hat{v}}_{s1}\\\nonumber
|\vect{\alpha}_{1}|^2 &= \frac{N}{2}(T_s+T_{ps})~\vect{\hat{v}}_{s1}^\dagger\matr{P}_{F}\matr{\Sigma}_{F}^{-1} \matr{P}_{F}^\dagger \vect{\hat{v}}_{s1} \\\nonumber
&= \frac{N}{2}(T_s+T_{ps})~\vect{\hat{v}}_{s1}^\dagger\matr{F}^{-1} \vect{\hat{v}}_{s1}\\\nonumber
\vect{\hat{\alpha}}_{1}  &= \frac{\vect{\alpha}_{1}}{|\vect{\alpha}_{1}|}=\frac{\matr{\Sigma}_{F}^{-1/2} \matr{P}_{F}^\dagger \vect{\hat{v}}_{s1}}{\left(\vect{\hat{v}}_{s1}^\dagger\matr{F}^{-1} \vect{\hat{v}}_{s1}\right)^{1/2}}.
\end{align}
Similarly,
\begin{align}
\label{eq:alpha2}
\vect{\alpha}_{2}&=\sqrt{\frac{N}{2}(T_s-T_{ps})}~\matr{\Sigma}_{F}^{-1/2} \matr{P}_{F}^\dagger \vect{\hat{v}}_{s2}\\\nonumber
|\vect{\alpha}_{2}|^2 &=  \frac{N}{2}(T_s-T_{ps})~\vect{\hat{v}}_{s2}^\dagger\matr{F}^{-1} \vect{\hat{v}}_{s2}\\\nonumber
\vect{\hat{\alpha}}_{2}  &= \frac{\vect{\alpha}_{2}}{|\vect{\alpha}_{2}|}=\frac{\matr{\Sigma}_{F}^{-1/2} \matr{P}_{F}^\dagger \vect{\hat{v}}_{s2}}{\left(\vect{\hat{v}}_{s2}^\dagger\matr{F}^{-1} \vect{\hat{v}}_{s2}\right)^{1/2}}.
\end{align}
Note that the vectors $\vect{\hat{v}}_{s1}$ and $\vect{\hat{v}}_{s2}$ are orthonormal and they are the two eigenvectors of $\matr{S}$ that span its image. However, $\vect{\alpha}_{1}$ and $\vect{\alpha}_{2}$ are not necessarily orthogonal:

\begin{align}
\label{eq:alpha1_dot_alpha2}
\vect{\alpha}_{1}^\dagger\vect{\alpha}_{2}&=\frac{N}{2}\sqrt{T_s^2-T_{ps}^2}~\vect{\hat{v}}_{s1}^\dagger\matr{F}^{-1} \vect{\hat{v}}_{s2}.
\end{align}

\subsubsection{\texorpdfstring{$\bm{\alpha_1},~\bm{\alpha_2}$}{alpha1, alpha2} orthogonal}
\label{subsec:orthogonal}

If $\vect{\hat{v}}_{s1}^\dagger\matr{F}^{-1} \vect{\hat{v}}_{s2}=0$ then $\vect{\alpha}_{1}$ and $\vect{\alpha}_{2}$ are orthogonal and the matrix $\matr{S''}$ in Equation~\ref{eq:D2_expanded} is already written in terms of its eigendecomposition. In this case, it is easy to check that $\vect{\hat{\alpha}}_{1}$ and $\vect{\hat{\alpha}}_{2}$ are the two eigenvectors of $\matr{S''}$ with non-zero eigenvalues. Their respective eigenvalues are $|\vect{\alpha}_{1}|^2$ and $|\vect{\alpha}_{2}|^2$. This also means that \jmp{$\lambda_1=|\vect{\alpha}_{1}|^2$} and \jmp{$\lambda_2=|\vect{\alpha}_{2}|^2$} are the two non-zero generalized eigenvalues of $\matr{S}$ and $\matr{F}$.

Note that since 

\begin{align}
\vect{\hat{\alpha}}_{a}^\dagger\matr{S''}\vect{\hat{\alpha}}_{a} = |\vect{\alpha}_{a}|^2, \hspace{0.2in} a=1, 2
\end{align}
then 

\begin{align}
\left(\matr{P}_{F}\matr{\Sigma}_{F}^{-1/2}\vect{\hat{\alpha}}_{a}\right)^\dagger\matr{S}\left(\matr{P}_{F}\matr{\Sigma}_{F}^{-1/2}\vect{\hat{\alpha}}_{a}\right) = \vect{b}_{a}^\dagger\matr{S}\vect{b}_{a} =|\vect{\alpha}_{a}|^2.
\end{align}

Similarly

\begin{align} \vect{b}_{a}^\dagger\matr{F}\vect{b}_{a} =1.
\end{align}

Thus, the two generalized eigenmodes (or KL beamformers) that contain signal power are

\begin{align}
\label{eq:klt_bf_multimode}
     \vect{b}_{a}=\matr{P}_{F}\matr{\Sigma}_{F}^{-1/2}\vect{\hat{\alpha}}_{a} =\frac{ \matr{F}^{-1}\vect{\hat{v}}_{sa}}{\left( \vect{\hat{v}}_{sa}^\dagger \matr{F}^{-1}\vect{\hat{v}}_{sa}\right)^{1/2}}, \hspace{0.2in} a=1,2
\end{align}
and the optimal signal-to-noise is

\begin{align}
\label{eq:snr_orthogonal}
    (S/N)_{KLT}=&|\vect{\alpha}_{1}|^2+|\vect{\alpha}_{2}|^2\\\nonumber
    =& \frac{N}{2}(T_s+T_{ps})~\vect{\hat{v}}_{s1}^\dagger\matr{F}^{-1} \vect{\hat{v}}_{s1} +
    \frac{N}{2}(T_s-T_{ps})~\vect{\hat{v}}_{s2}^\dagger\matr{F}^{-1} \vect{\hat{v}}_{s2}
\end{align}
\subsubsection{\texorpdfstring{$\bm{\alpha_1},~\bm{\alpha_2}$}{alpha1, alpha2} not orthogonal}
\label{subsec:not_orthogonal}

If $\vect{\hat{v}}_{s1}^\dagger\matr{F}^{-1} \vect{\hat{v}}_{s2}\neq 0$ then $\vect{\alpha}_{1}$ and $\vect{\alpha}_{2}$ are not orthogonal and they are not eigenvectors of $\matr{S''}$.

Even without the eigenvalue decomposition of $\matr{S''}$ we can read $(S/N)_{KLT}$ from Equation~\ref{eq:D2_expanded}. By using Equation~\ref{eq:S2} we have

\begin{align}
\label{eq:snr_nonorthogonal}
    (S/N)_{KLT}=&\text{Tr}(\matr{\Lambda})=\text{Tr}(\matr{S''})\\\nonumber
    =&|\vect{\alpha}_{1}|^2+|\vect{\alpha}_{2}|^2\\\nonumber
\end{align}
so the total $S/N$ in this case will be the same as in Equation~\ref{eq:snr_orthogonal}.

In order to know the individual $S/N$ contribution of each signal mode we need to find the eigenvalue decomposition of $\matr{S''}$, or at least the two eigenvectors that span its image.

From inspection, the two eigenvectors will have the form

\begin{align}
\label{eq:nonorthogonal_evectors}
    \vect{\beta}_{\pm} = \vect{\alpha}_{1}+\kappa_{\pm}e^{-i\phi}\vect{\alpha}_{2}
\end{align}
where

\begin{align}
e^{-i\phi} =  \frac{\vect{\alpha}_{2}^\dagger\vect{\alpha}_{1}}{|\vect{\alpha}_{2}^\dagger\vect{\alpha}_{1}|}
\end{align}
and

\begin{align}
\kappa_{\pm} =  \frac{-|\vect{\alpha}_{1}|^2+|\vect{\alpha}_{2}|^2 \pm \sqrt{\left(|\vect{\alpha}_{1}|^2-|\vect{\alpha}_{2}|^2\right)^2+4|\vect{\alpha}_{1}^\dagger\vect{\alpha}_{2}|^2}}{2|\vect{\alpha}_{1}^\dagger\vect{\alpha}_{2}|}.
\end{align}

The corresponding eigenvalues are

\begin{align}
\lambda_{\pm} =  \frac{1}{2}\left[|\vect{\alpha}_{1}|^2+|\vect{\alpha}_{2}|^2 \pm \sqrt{\left(|\vect{\alpha}_{1}|^2-|\vect{\alpha}_{2}|^2\right)^2+4|\vect{\alpha}_{1}^\dagger\vect{\alpha}_{2}|^2}\right]
\end{align}
which confirms Equation~\ref{eq:snr_nonorthogonal}.

Thus, the two generalized eigenmodes (or KL beamformers) that contain signal power are

\begin{align}
\label{eq:klt_bf_multimode_nonorthogonal}
     \vect{b}_{\pm}&=\matr{P}_{F}\matr{\Sigma}_{F}^{-1/2}\vect{\hat{\beta}}_{\pm}\\
     &=\frac{1}{|\vect{\beta}_{\pm}|} \matr{F}^{-1}\left[\sqrt{\frac{N}{2}(T_s+T_{ps})}~ \vect{\hat{v}}_{s1}+\sqrt{\frac{N}{2}(T_s-T_{ps})}~\kappa_{\pm}e^{-i\phi} \vect{\hat{v}}_{s2}\right].
\end{align}

\jmp{
\subsubsection{Constraints on optimal total $S/N$}
\label{subsec:total_sn_constraints}

When the source is completely polarized its covariance matrix $\matr{S}$ consists of a single mode and the KL beamformer optimizes the total $S/N$. In the more general case of a partially polarized source, $\matr{S}$ consists of two independent modes. As explained in Section~\ref{sec:klt}, the KLT optimizes the sum of per-mode $S/N$ values (Equation~\ref{eq:klt_snr_tot}), but not necessarily the total $S/N$ (Equation~\ref{eq:trace_ratio_multimode}). \fix{However, as we will show in this section, the KLT provides bounds for the optimal total $S/N$, and in special cases is equivalent to the solution which optimizes the total $S/N$.}

\fix{If the matrix $\tilde{\matr{B}}=\left[\vect{b}_1,~\vect{b}_2\right]$ represents two orthonormal beamformers (i.e. $\vect{b}_k^\dagger \vect{b}_l = \delta_{kl}$) acting on the full polarization data, then the total $S/N$ becomes
\begin{align}
    \label{eq:SN_tot_unpol_inequality0}
   \frac{\text{Tr} \left(\matr{\tilde{B}}^\dagger \matr{S} \matr{\tilde{B}}\right)}{\text{Tr} \left(\matr{\tilde{B}}^\dagger \matr{F} \matr{\tilde{B}}\right)} =\frac{\sum_k\vect{b}_k^\dagger \matr{S} \vect{b}_k}{\sum_k\vect{b}_k^\dagger \matr{F} \vect{b}_k} 
\end{align}
Since the total number of modes (described by index $k$) is 2, we can expand Equation \ref{eq:SN_tot_unpol_inequality0} as
\begin{align}
\label{eq:SN_tot_unpol_inequality1}
    \frac{\text{Tr} \left(\matr{\tilde{B}}^\dagger \matr{S} \matr{\tilde{B}}\right)}{\text{Tr} \left(\matr{\tilde{B}}^\dagger \matr{F} \matr{\tilde{B}}\right)} &= \frac{\vect{b}_1^\dagger \matr{F} \vect{b}_1}{\sum_k\vect{b}_k^\dagger \matr{F} \vect{b}_k}\cdot\frac{\vect{b}_1^\dagger \matr{S} \vect{b}_1}{\vect{b}_1^\dagger \matr{F} \vect{b}_1} + \frac{\vect{b}_2^\dagger \matr{F} \vect{b}_2}{\sum_k\vect{b}_k^\dagger \matr{F} \vect{b}_k}\cdot\frac{\vect{b}_2^\dagger \matr{S} \vect{b}_2}{\vect{b}_2^\dagger \matr{F} \vect{b}_2} \\
   &\leq \text{Max}\left\{\frac{\vect{b}_1^\dagger \matr{S} \vect{b}_1}{\vect{b}_1^\dagger \matr{F} \vect{b}_1}, \frac{\vect{b}_2^\dagger \matr{S} \vect{b}_2}{\vect{b}_2^\dagger \matr{F} \vect{b}_2}\right\}\nonumber
\end{align}
where the first line follows from the fact that $\matr{F}$ is positive definite ($\vect{b}_2^\dagger \matr{F} \vect{b}_2$ and $\vect{b}_1^\dagger \matr{F} \vect{b}_1$ are positive) and the Max function in the second line denotes the maximum of the two single-mode $S/N$ arguments. \jmp{Also, from Section~\ref{sec:klt_bf} we know that for any unit-norm beamformer $\vect{b}$ acting on $\matr{S}$ and $\matr{F}$ we must have

\begin{align}
\label{eq:rayleigh_inequality}
  \frac{\vect{b}^\dagger \matr{S} \vect{b}}{\vect{b}^\dagger \matr{F} \vect{b}} \leq \lambda_*
\end{align}

where $\lambda_*$ is the largest generalized eigenvalue of the pair $\matr{S} -\matr{F}$.} This implies that Equation \ref{eq:SN_tot_unpol_inequality0} must follow the inequality:
\begin{align}
\label{eq:SN_tot_unpol_inequality}
\frac{\text{Tr} \left(\matr{\tilde{B}}^\dagger \matr{S} \matr{\tilde{B}}\right)}{\text{Tr} \left(\matr{\tilde{B}}^\dagger \matr{F} \matr{\tilde{B}}\right)}& \leq \lambda_*.
\end{align}
Equipped with Equation \ref{eq:SN_tot_unpol_inequality}, let's now  consider the special case where} the source is unpolarized ($Q_s=U_s=V_s=0$). From Equations~\ref{eq:S_pol} and \ref{eq:S_pol1}, $\matr{S}$ is rank-2 with two identical non-zero eigenvalues and block diagonal with

\begin{align}
\matr{S}^{XX}=\frac{NT_s}{2}\vect{\hat{v}}_{s}^X\vect{\hat{v}}_{s}^{X\dagger}, \hspace{0.1in} \matr{S}^{YY}=\frac{NT_s}{2}\vect{\hat{v}}_{s}^Y\vect{\hat{v}}_{s}^{Y\dagger}, \hspace{0.1in}\matr{S}^{XY}=\matr{S}^{YX}=\matr{0}.
\end{align}
Furthermore, we can always choose the indexing of our dual-polarization antenna array such that $\vect{\hat{v}}_{s}^X=\vect{\hat{v}}_{s}^Y$ (Section~\ref{sec:polarization}), so $\matr{S}^{XX}=\matr{S}^{YY}$. In addition, if the sky contamination is also unpolarized and the receivers are identical without cross-polarization noise terms, then $\matr{F}$ will also have a similar block diagonal structure with $\matr{F}^{XX}=\matr{F}^{YY}$. The two identical blocks of the system imply that the pair $\matr{S} -\matr{F}$ will have two identical non-zero generalized eigenvalues $(\lambda_*,~\lambda_*)$, with $\lambda_*$ being the only non-zero generalized eigenvalue of the pair $\matr{S}^{XX}- \matr{F}^{XX}$, that is \fix{(Section~\ref{sec:klt_bf})}

\begin{align}
\label{eq:rayleigh_unpolarized}
  \underset{\fix{\vect{b}^{X\dagger}\vect{b}^X=1}}{\text{Max}}\left\{\frac{\vect{b}^{X\dagger} \matr{S}^{XX} \vect{b}^{X}}{\vect{b}^{X\dagger} \matr{F}^{XX} \vect{b}^{X}}\right\} &=\lambda_*
\end{align}
\fix{where $\vect{b}^{X}$ represents a single-polarization beamformer. Note that we use a normalization constraint (unit vector norm) that is different to the constraint in Equation~\ref{eq:klt_bf_opt1} (unit denominator), but that leaves the $S/N$ and the optimization procedure unchanged.} Also, from Equation~\ref{eq:klt_bf} we know that the vector
\begin{align}
\label{eq:klt_bf_single_pol}
     \vect{b}_*^X=\frac{ \matr{F}^{-1}\vect{\hat{v}}_{s}^X}{\left( \vect{\hat{v}}_{s}^{X\dagger} \matr{F}^{-2}\vect{\hat{v}}_{s}^X\right)^{1/2}}
\end{align}
is the optimal solution for Equation~\ref{eq:rayleigh_unpolarized}, \fix{where the different value of the denominator in this case is determined by the unit vector norm constraint}. This means that if we set 

\begin{equation}
\label{eq:KL_unpolarized_soln}
\vect{b}_{1*} = \begin{bmatrix}
\vect{b}_*^X \\
\vect{0}
\end{bmatrix}, ~ \vect{b}_{2*} = \begin{bmatrix}
\vect{0} \\
\vect{b}_*^X
\end{bmatrix} \hspace{0.2in} \rightarrow \hspace{0.2in} \tilde{\matr{B}}_*=\left[\vect{b}_{1*},~\vect{b}_{2*}\right]
\end{equation}

we have

\begin{align}
\label{eq:SN_tot_unpol}
   \frac{\text{Tr} \left(\matr{\tilde{B}}_*^\dagger \matr{S} \matr{\tilde{B}}_*\right)}{\text{Tr} \left(\matr{\tilde{B}}_*^\dagger \matr{F} \matr{\tilde{B}}_*\right)}&=\frac{2~ \vect{b}_*^{X\dagger} \matr{S}^{XX} \vect{b}_*^X}{2~\vect{b}_*^{X\dagger} \matr{F}^{XX} \vect{b}_*^X}= \lambda_*.
\end{align}

It follows from Equations~\ref{eq:SN_tot_unpol_inequality} and \ref{eq:SN_tot_unpol} that, in the unpolarized limit, \fix{$\tilde{\matr{B}}_*$ as given by Equations \ref{eq:klt_bf_single_pol} and \ref{eq:KL_unpolarized_soln} provides the optimal total $S/N$ and}

\begin{align}
\label{eq:trace_ratio_multimode_unpol}
      \underset{\matr{\tilde{B}}^\dagger \matr{\tilde{B}} = \matr{I}_2}{\text{Max}}\left\{\frac{\text{Tr} \left(\matr{\tilde{B}}^\dagger \matr{S} \matr{\tilde{B}}\right)}{\text{Tr} \left(\matr{\tilde{B}}^\dagger \matr{F} \matr{\tilde{B}}\right)}\right\}_{unpolarized} = \lambda_*.
\end{align}
\fix{Thus, the unpolarized case is a rank-2 trace ratio problem that can be solved by finding the KL beamformer for a single polarization (a simpler, rank-1 problem).}

\fix{The previous statement is not true in general. However,} from the work of \cite{doi:10.1137/120864799} on the localization of the optimum for the general rank-$k>1$ trace ratio problem it follows that, in the most general partially polarized case \fix{(including partially polarized source and contamination, as well as potential co- and cross-polarization noise couplings)}, the non-zero generalized eigenvalues will provide bounds for the optimal total $S/N$, that is

\begin{align}
\label{eq:trace_ratio_multimode_inequality}
     \lambda_2  \leq \underset{\matr{\tilde{B}}^\dagger \matr{\tilde{B}} = \matr{I}_2}{\text{Max}}\left\{\frac{\text{Tr} \left(\matr{\tilde{B}}^\dagger \matr{S} \matr{\tilde{B}}\right)}{\text{Tr} \left(\matr{\tilde{B}}^\dagger \matr{F} \matr{\tilde{B}}\right)}\right\} \leq \lambda_1
\end{align}
where $\lambda_1\geq \lambda_2$ are the two non-zero generalized eigenvalues of $\matr{S}$ and $\matr{F}$.
 }

\fix{
\subsection{Visibility space}

Sometimes the interferometric data available is not in the form of baseband voltages but as a per-frequency estimate $\tilde{\matr{D}}$ of the $2N \times 2N$ visibility matrix (Equation~\ref{eq:data_covariance}). This means that $\tilde{D}_{ij}$ represents an estimate of the correlation between antenna elements $i$ and $j$ in the form of Equation~\ref{eq:diagonal_loading}, and that $\left \langle \tilde{\matr{D}} \right \rangle=\matr{D}$.

In this case, one can form beamformed visibility beams by taking linear combinations of visibilities 
\begin{equation}
    y = \sum_{ij} w_{ij}^* \tilde{D}_{ij} = \vect{w}^\dagger\text{vect}(\tilde{\matr{D}}) = \text{Tr}(\matr{W}^\dagger\tilde{\matr{D}})
\end{equation}

where $\text{vect}(\tilde{\matr{D}})$ is the column-vectorized version of $\tilde{\matr{D}}$ formed by stacking its columns, $\vect{w}$ is the length-$(2N)^2$ visibility beamforming vector, and $\matr{W}$ is its corresponding $2N \times 2N$ matrix representation. 

Voltage beamformers discussed in previous sections are a sub-class of visibility beamformers. To see this, note that if $\vect{b}^\dagger\vect{d}(t_m)$ is a beamformed voltage beam of the $m$-th voltage data sample, then the corresponding visibility beam is
\begin{align}
    y_b &=  \frac{1}{M} \sum_m \left |\vect{b}^\dagger\vect{d}(t_m)\right |^2 = \vect{b}^\dagger \tilde{\matr{D}}\vect{b} = \text{Tr}(\matr{W}_b^\dagger\tilde{\matr{D}})
\end{align}
where
\begin{equation}
    \matr{W}_b = \vect{b}\vect{b}^\dagger.
\end{equation}
Thus, voltage beams are the sub-class of visibility beams that are ``factorizable".

In the case of visibility beamformers the $S/N$ is typically measured as the ratio of the average signal power 

\begin{equation}
    \bar{y}_S = \left \langle y_S \right \rangle =\vect{w}^\dagger\text{vect}(\matr{S})
\end{equation}

to the root mean square noise and sky interference power \citep{Interferometry_Thompson}, $\Delta y_F$, where

\begin{equation}
    (\Delta y_F)^2 = \left \langle |y_F|^2 \right \rangle -|\left \langle y_F \right \rangle |^2 =\vect{w}^\dagger\bm{\mathcal{F}}\vect{w}
\end{equation}

and

\begin{equation}
\label{eq:mathcal_F}
    \bm{\mathcal{F}} = \left \langle \text{vect}(\tilde{\matr{F}}) \text{vect}(\tilde{\matr{F}})^\dagger \right \rangle - \text{vect}(\matr{F})\text{vect}(\matr{F})^\dagger
\end{equation}

is the covariance matrix of the noise and sky interference visibilities.

This means that the optimal visibility beamforming weight vector $\vect{w}$ can be found by maximizing $|\bar{y}_S|^2/(\Delta y_F)^2$ or, equivalently, by solving the optimization problem

\begin{align}
\label{eq:klt_vis}
    \underset{\vect{w}}{\text{Max}}\left\{\frac{\vect{w}^\dagger \bm{\mathcal{S}} \vect{w}}{\vect{w}^\dagger \bm{\mathcal{F}} \vect{w}}\right\}.
\end{align}

where we defined the rank-1 matrix $\bm{\mathcal{S}} = \text{vect}(\matr{S})\text{vect}(\matr{S})^\dagger$.

From Section~\ref{sec:klt_bf}, we know that the KLT provides the solution for Equation~\ref{eq:klt_vis} and that the  optimal visibility KL beamformer is

\begin{align}
\label{eq:klt_bf_vis}
     \vect{w}\propto \bm{\mathcal{F}}^{-1}\text{vect}(\matr{S}).
\end{align}

Note that by the same argument presented in Equation \ref{eq:diagonal_loading}, $N^2$ time samples are now needed to enforce that F is well-measured, which for CHIME would require seconds of data.


In the case that the noise and the sky interference signals are Gaussian, the matrix $\bm{\mathcal{F}}$ takes a simple form \citep{2015_Masui}

\begin{align}
\label{eq:mathcal_F_gaussian}
     \matr{\mathcal{F}}_{(ij)(kl)} = \frac{F_{ik}F_{jl}^*}{M}, \hspace{0.4in} \left(\matr{\mathcal{F}}^{-1}\right)_{(ij)(kl)} = M\refereejmp{\left(F^{-1}\right)_{ik}\left(F^{-1}\right)_{jl}^*}
\end{align}

and the optimal visibility beamformer is equivalent to the solution presented in Equation~32 of \cite{Masui_2017}. In general, however, given that many sources of RFI are non-Gaussian \citep{taylor_2019}, Equation~\ref{eq:mathcal_F_gaussian}- will not hold, and the full noise and the sky interference covariance given by Equation \ref{eq:mathcal_F} will need to be estimated from the visibility data. }

\bibliography{references}{}

@article{chatterjee2017direct,
	title        = {{A direct localization of a fast radio burst and its host}},
	author       = {{Chatterjee}, S. and {Law}, C.~J. and {Wharton}, R.~S. and {Burke-Spolaor}, S. and {Hessels}, J.~W.~T. and {Bower}, G.~C. and {Cordes}, J.~M. and {Tendulkar}, S.~P. and {Bassa}, C.~G. and {Demorest}, P. and {Butler}, B.~J. and {Seymour}, A. and {Scholz}, P. and {Abruzzo}, M.~W. and {Bogdanov}, S. and {Kaspi}, V.~M. and {Keimpema}, A. and {Lazio}, T.~J.~W. and {Marcote}, B. and {McLaughlin}, M.~A. and {Paragi}, Z. and {Ransom}, S.~M. and {Rupen}, M. and {Spitler}, L.~G. and {van Langevelde}, H.~J.},
	year         = 2017,
	month        = jan,
	journal      = {\nat},
	volume       = 541,
	number       = 7635,
	pages        = {58--61},
	doi          = {10.1038/nature20797},
	archiveprefix = {arXiv},
	eprint       = {1701.01098},
	primaryclass = {astro-ph.HE},
	adsurl       = {https://ui.adsabs.harvard.edu/abs/2017Natur.541...58C}
}

@article{michilli2020analysis,
	title        = {{An analysis pipeline for CHIME/FRB full-array baseband data}},
	author       = {{Michilli}, D. and {Masui}, K.~W. and {Mckinven}, R. and {Cubranic}, D. and {Bruneault}, M. and {Brar}, C. and {Patel}, C. and {Boyle}, P.~J. and {Stairs}, I.~H. and {Renard}, A. and {Bandura}, K. and {Berger}, S. and {Breitman}, D. and {Cassanelli}, T. and {Dobbs}, M. and {Kaspi}, V.~M. and {Leung}, C. and {Mena-Parra}, J. and {Pleunis}, Z. and {Russell}, L. and {Scholz}, P. and {Siegel}, S.~R. and {Tendulkar}, S.~P. and {Vand erlinde}, K.},
	year         = 2020,
	month        = oct,
	journal      = {arXiv e-prints},
	pages        = {arXiv:2010.06748},
	eid          = {arXiv:2010.06748},
	archiveprefix = {arXiv},
	eprint       = {2010.06748},
	primaryclass = {astro-ph.HE},
	adsurl       = {https://ui.adsabs.harvard.edu/abs/2020arXiv201006748M}
}

@ARTICLE{1444313,
  author={North, D.O.},
  journal={Proceedings of the IEEE}, 
  title={An Analysis of the factors which determine signal/noise discrimination in pulsed-carrier systems}, 
  year={1963},
  volume={51},
  number={7},
  pages={1016-1027},
  doi={10.1109/PROC.1963.2383}}

@ARTICLE{Chatterjee_2009,
       author = {{Chatterjee}, S. and {Brisken}, W.~F. and {Vlemmings}, W.~H.~T. and {Goss}, W.~M. and {Lazio}, T.~J.~W. and {Cordes}, J.~M. and {Thorsett}, S.~E. and {Fomalont}, E.~B. and {Lyne}, A.~G. and {Kramer}, M.},
        title = "{Precision Astrometry with the Very Long Baseline Array: Parallaxes and Proper Motions for 14 Pulsars}",
      journal = {\apj},
         year = 2009,
        month = jun,
       volume = {698},
       number = {1},
        pages = {250-265},
          doi = {10.1088/0004-637X/698/1/250},
archivePrefix = {arXiv},
       eprint = {0901.1436},
 primaryClass = {astro-ph.SR},
       adsurl = {https://ui.adsabs.harvard.edu/abs/2009ApJ...698..250C}
}

@ARTICLE{2019arXiv190311240G,
       author = {{Ghojogh}, Benyamin and {Karray}, Fakhri and {Crowley}, Mark},
        title = "{Eigenvalue and Generalized Eigenvalue Problems: Tutorial}",
      journal = {arXiv e-prints},
         year = 2019,
        month = mar,
          eid = {arXiv:1903.11240},
        pages = {arXiv:1903.11240},
          doi = {10.48550/arXiv.1903.11240},
archivePrefix = {arXiv},
       eprint = {1903.11240},
 primaryClass = {stat.ML},
       adsurl = {https://ui.adsabs.harvard.edu/abs/2019arXiv190311240G}
}

@article{Cassanelli_2024,
   title={A fast radio burst localized at detection to an edge-on galaxy using very-long-baseline interferometry},
   volume={8},
   ISSN={2397-3366},
   url={http://dx.doi.org/10.1038/s41550-024-02357-x},
   DOI={10.1038/s41550-024-02357-x},
   number={11},
   journal={Nature Astronomy},
   publisher={Springer Science and Business Media LLC},
   author={Cassanelli, Tomas and Leung, Calvin and Sanghavi, Pranav and Mena-Parra, Juan and Cary, Savannah and Mckinven, Ryan and Bhardwaj, Mohit and Masui, Kiyoshi W. and Michilli, Daniele and Bandura, Kevin and Chatterjee, Shami and Peterson, Jeffrey B. and Kaczmarek, Jane and Rahman, Mubdi and Shin, Kaitlyn and Vanderlinde, Keith and Berger, Sabrina and Brar, Charanjot and Boyle, P. J. and Breitman, Daniela and Chawla, Pragya and Curtin, Alice P. and Dobbs, Matt and Dong, Fengqiu Adam and Fonseca, Emmanuel and Gaensler, B. M. and Ibik, Adaeze and Kaspi, Victoria M. and Khairy, Kholoud and Lanman, Adam E. and Lazda, Mattias and Lin, Hsiu-Hsien and Luo, Jing and Meyers, Bradley W. and Milutinovic, Nikola and Ng, Cherry and Noble, Gavin and Pearlman, Aaron B. and Pen, Ue-Li and Petroff, Emily and Pleunis, Ziggy and Quine, Brendan and Rafiei-Ravandi, Masoud and Renard, Andre and Sand, Ketan R. and Schoen, Eve and Scholz, Paul and Smith, Kendrick M. and Stairs, Ingrid and Tendulkar, Shriharsh P.},
   year={2024},
   month=sep, pages={1429–1442} }

@ARTICLE{Outriggers_2025,
       author = {{CHIME/FRB Collaboration} and {Amiri}, Mandana and {Andersen}, Bridget C. and {Andrew}, Shion and {Bandura}, Kevin and {Bhardwaj}, Mohit and {Bhopi}, Kalyani and {Bidula}, Vadym and {Boyle}, P.~J. and {Brar}, Charanjot and {Carlson}, Mark and {Cassanelli}, Tomas and {Cassity}, Alyssa and {Chatterjee}, Shami and {Cliche}, Jean-Fran{\c{c}}ois and {Curtin}, Alice P. and {Darlinger}, Rachel and {DeBoer}, David R. and {Dobbs}, Matt and {Dong}, Fengqiu Adam and {Eadie}, Gwendolyn and {Fonseca}, Emmanuel and {Gaensler}, B.~M. and {Gusinskaia}, Nina and {Halpern}, Mark and {Hendricksen}, Ian and {Hessels}, Jason and {Joseph}, Ronniy C. and {Kaczmarek}, Jane and {Kaspi}, Victoria M. and {Khairy}, Kholoud and {Landecker}, T.~L. and {Lanman}, Adam E. and {Kit Lau}, Albert Wai and {Lazda}, Mattias and {Leung}, Calvin and {Main}, Robert A. and {Masui}, Kiyoshi W. and {Mckinven}, Ryan and {Mena-Parra}, Juan and {Meyers}, Bradley W. and {Michilli}, Daniele and {Milutinovic}, Nikola and {Nimmo}, Kenzie and {Noble}, Gavin and {Pandhi}, Ayush and {Pearlman}, Aaron B. and {Peterson}, Jeffrey B. and {Petroff}, Emily and {Pleunis}, Ziggy and {Pollak}, Alexander W. and {Rafiei-Ravandi}, Masoud and {Renard}, Andre and {Sammons}, Mawson W. and {Sand}, Ketan R. and {Sanghavi}, Pranav and {Scholz}, Paul and {Shah}, Vishwangi and {Shin}, Kaitlyn and {Siegel}, Seth R. and {Siemion}, Andrew and {Sievers}, Jonathan L. and {Smith}, Kendrick and {Spear}, David and {Stairs}, Ingrid and {Vanderlinde}, Keith and {Wang}, Haochen and {Willis}, Jacob P. and {Zegmott}, Tarik J.},
        title = "{CHIME/FRB Outriggers: Design Overview}",
      journal = {arXiv e-prints},
         year = 2025,
        month = apr,
          eid = {arXiv:2504.05192},
        pages = {arXiv:2504.05192},
          doi = {10.48550/arXiv.2504.05192},
archivePrefix = {arXiv},
       eprint = {2504.05192},
 primaryClass = {astro-ph.HE},
       adsurl = {https://ui.adsabs.harvard.edu/abs/2025arXiv250405192C}
}

@ARTICLE{BURSTT_2022,
       author = {{Lin}, Hsiu-Hsien and {Lin}, Kai-yang and {Li}, Chao-Te and {Tseng}, Yao-Huan and {Jiang}, Homin and {Wang}, Jen-Hung and {Cheng}, Jen-Chieh and {Pen}, Ue-Li and {Chen}, Ming-Tang and {Chen}, Pisin and {Chen}, Yaocheng and {Goto}, Tomotsugu and {Hashimoto}, Tetsuya and {Hwang}, Yuh-Jing and {King}, Sun-Kun and {Kubo}, Derek and {Kuo}, Chung-Yun and {Mills}, Adam and {Nam}, Jiwoo and {Oshiro}, Peter and {Shen}, Chang-Shao and {Tseng}, Hsien-Chun and {Wang}, Shih-Hao and {Wu}, Vigo Feng-Shun and {Bower}, Geoffrey and {Chang}, Shu-Hao and {Chen}, Pai-An and {Chen}, Ying-Chih and {Chiang}, Yi-Kuan and {Fedynitch}, Anatoli and {Gusinskaia}, Nina and {Ho}, Simon C. -C. and {Hsiao}, Tiger Y. -Y. and {Hu}, Chin-Ping and {Huang}, Yau De and {J{\'a}uregui Garc{\'\i}a}, Jos{\'e} Miguel and {Kim}, Seong Jin and {Kuo}, Cheng-Yu and {Ling}, Decmend Fang-Jie and {On}, Alvina Y.~L. and {Peterson}, Jeffrey B. and {R. Raquel}, Bjorn Jasper and {Su}, Shih-Chieh and {Uno}, Yuri and {Wu}, Cossas K. -W. and {Yamasaki}, Shotaro and {Zhu}, Hong-Ming},
        title = "{BURSTT: Bustling Universe Radio Survey Telescope in Taiwan}",
      journal = {\pasp},
         year = 2022,
        month = sep,
       volume = {134},
       number = {1039},
          eid = {094106},
        pages = {094106},
          doi = {10.1088/1538-3873/ac8f71},
archivePrefix = {arXiv},
       eprint = {2206.08983},
 primaryClass = {astro-ph.IM},
       adsurl = {https://ui.adsabs.harvard.edu/abs/2022PASP..134i4106L}
}

@ARTICLE{taylor_2019,
       author = {{Taylor}, Jacob and {Denman}, Nolan and {Bandura}, Kevin and {Berger}, Philippe and {Masui}, Kiyoshi and {Renard}, Andre and {Tretyakov}, Ian and {Vanderlinde}, Keith},
        title = "{Spectral Kurtosis-Based RFI Mitigation for CHIME}",
      journal = {Journal of Astronomical Instrumentation},
         year = 2019,
        month = jan,
       volume = {8},
       number = {1},
          eid = {1940004},
        pages = {1940004},
          doi = {10.1142/S225117171940004X},
archivePrefix = {arXiv},
       eprint = {1808.10365},
 primaryClass = {astro-ph.IM},
       adsurl = {https://ui.adsabs.harvard.edu/abs/2019JAI.....840004T}
}

@ARTICLE{Masoud_2023,
       author = {{Rafiei-Ravandi}, Masoud and {Smith}, Kendrick M.},
        title = "{Mitigating Radio Frequency Interference in CHIME/FRB Real-time Intensity Data}",
      journal = {\apjs},
         year = 2023,
        month = apr,
       volume = {265},
       number = {2},
          eid = {62},
        pages = {62},
          doi = {10.3847/1538-4365/acc252},
archivePrefix = {arXiv},
       eprint = {2206.07292},
 primaryClass = {astro-ph.IM},
       adsurl = {https://ui.adsabs.harvard.edu/abs/2023ApJS..265...62R}
}

@ARTICLE{Shaw_2015,
       author = {{Shaw}, J. Richard and {Sigurdson}, Kris and {Sitwell}, Michael and {Stebbins}, Albert and {Pen}, Ue-Li},
        title = "{Coaxing cosmic 21 cm fluctuations from the polarized sky using m -mode analysis}",
      journal = {\prd},
         year = 2015,
        month = apr,
       volume = {91},
       number = {8},
          eid = {083514},
        pages = {083514},
          doi = {10.1103/PhysRevD.91.083514},
archivePrefix = {arXiv},
       eprint = {1401.2095},
 primaryClass = {astro-ph.CO},
       adsurl = {https://ui.adsabs.harvard.edu/abs/2015PhRvD..91h3514S}
}

@INPROCEEDINGS{Vanderlinde_2019,
       author = {{Vanderlinde}, Keith and {Liu}, Adrian and {Gaensler}, Bryan and {Bond}, Dick and {Hinshaw}, Gary and {Ng}, Cherry and {Chiang}, Cynthia and {Stairs}, Ingrid and {Brown}, Jo-Anne and {Sievers}, Jonathan and {Mena}, Juan and {Smith}, Kendrick and {Bandura}, Kevin and {Masui}, Kiyoshi and {Spekkens}, Kristine and {Belostotski}, Leo and {Dobbs}, Matt and {Turok}, Neil and {Boyle}, Patrick and {Rupen}, Michael and {Landecker}, Tom and {Pen}, Ue-Li and {Kaspi}, Victoria},
        title = "{The Canadian Hydrogen Observatory and Radio-transient Detector (CHORD)}",
    booktitle = {Canadian Long Range Plan for Astronomy and Astrophysics White Papers},
         year = 2019,
       volume = {2020},
        month = oct,
          eid = {28},
        pages = {28},
          doi = {10.5281/zenodo.3765414},
archivePrefix = {arXiv},
       eprint = {1911.01777},
 primaryClass = {astro-ph.IM},
       adsurl = {https://ui.adsabs.harvard.edu/abs/2019clrp.2020...28V}
}

@INPROCEEDINGS{Raybole_2019,
  author={Raybole, Pravin Ashok and Sureshkumar, S and Ankur, Ankur and Rai, Sanjeet Kumar and Bhujbal, Shrikant and Chaudhary, Lalit},
  booktitle={2019 IEEE MTT-S International Microwave and RF Conference (IMARC)}, 
  title={RFI shielded enclosure for radio telescope receiver electronics}, 
  year={2019},
  volume={},
  number={},
  pages={1-5},
  doi={10.1109/IMaRC45935.2019.9118755}}

@ARTICLE{Offringa_2013,
       author = {{Offringa}, A.~R. and {de Bruyn}, A.~G. and {Zaroubi}, S. and {van Diepen}, G. and {Martinez-Ruby}, O. and {Labropoulos}, P. and {Brentjens}, M.~A. and {Ciardi}, B. and {Daiboo}, S. and {Harker}, G. and {Jeli{\'c}}, V. and {Kazemi}, S. and {Koopmans}, L.~V.~E. and {Mellema}, G. and {Pandey}, V.~N. and {Pizzo}, R.~F. and {Schaye}, J. and {Vedantham}, H. and {Veligatla}, V. and {Wijnholds}, S.~J. and {Yatawatta}, S. and {Zarka}, P. and {Alexov}, A. and {Anderson}, J. and {Asgekar}, A. and {Avruch}, M. and {Beck}, R. and {Bell}, M. and {Bell}, M.~R. and {Bentum}, M. and {Bernardi}, G. and {Best}, P. and {Birzan}, L. and {Bonafede}, A. and {Breitling}, F. and {Broderick}, J.~W. and {Br{\"u}ggen}, M. and {Butcher}, H. and {Conway}, J. and {de Vos}, M. and {Dettmar}, R.~J. and {Eisloeffel}, J. and {Falcke}, H. and {Fender}, R. and {Frieswijk}, W. and {Gerbers}, M. and {Griessmeier}, J.~M. and {Gunst}, A.~W. and {Hassall}, T.~E. and {Heald}, G. and {Hessels}, J. and {Hoeft}, M. and {Horneffer}, A. and {Karastergiou}, A. and {Kondratiev}, V. and {Koopman}, Y. and {Kuniyoshi}, M. and {Kuper}, G. and {Maat}, P. and {Mann}, G. and {McKean}, J. and {Meulman}, H. and {Mevius}, M. and {Mol}, J.~D. and {Nijboer}, R. and {Noordam}, J. and {Norden}, M. and {Paas}, H. and {Pandey}, M. and {Pizzo}, R. and {Polatidis}, A. and {Rafferty}, D. and {Rawlings}, S. and {Reich}, W. and {R{\"o}ttgering}, H.~J.~A. and {Schoenmakers}, A.~P. and {Sluman}, J. and {Smirnov}, O. and {Sobey}, C. and {Stappers}, B. and {Steinmetz}, M. and {Swinbank}, J. and {Tagger}, M. and {Tang}, Y. and {Tasse}, C. and {van Ardenne}, A. and {van Cappellen}, W. and {van Duin}, A.~P. and {van Haarlem}, M. and {van Leeuwen}, J. and {van Weeren}, R.~J. and {Vermeulen}, R. and {Vocks}, C. and {Wijers}, R.~A.~M.~J. and {Wise}, M. and {Wucknitz}, O.},
        title = "{The LOFAR radio environment}",
      journal = {\aap},
         year = 2013,
        month = jan,
       volume = {549},
          eid = {A11},
        pages = {A11},
          doi = {10.1051/0004-6361/201220293},
archivePrefix = {arXiv},
       eprint = {1210.0393},
 primaryClass = {astro-ph.IM},
       adsurl = {https://ui.adsabs.harvard.edu/abs/2013A&A...549A..11O}
}

@ARTICLE{Chen_2025,
       author = {{Chen}, Kai-Feng and {Wilensky}, Michael J. and {Liu}, Adrian and {Dillon}, Joshua S. and {Hewitt}, Jacqueline N. and {Adams}, Tyrone and {Aguirre}, James E. and {Baartman}, Rushelle and {Beardsley}, Adam P. and {Berkhout}, Lindsay M. and {Bernardi}, Gianni and {Billings}, Tashalee S. and {Bowman}, Judd D. and {Bull}, Philip and {Burba}, Jacob and {Byrne}, Ruby and {Carey}, Steven and {Choudhuri}, Samir and {Cox}, Tyler and {DeBoer}, David. R. and {Dexter}, Matt and {Eksteen}, Nico and {Ely}, John and {Ewall-Wice}, Aaron and {Furlanetto}, Steven R. and {Gale-Sides}, Kingsley and {Garsden}, Hugh and {Gehlot}, Bharat Kumar and {Gorce}, Ad{\'e}lie and {Gorthi}, Deepthi and {Halday}, Ziyaad and {Hazelton}, Bryna J. and {Hickish}, Jack and {Jacobs}, Daniel C. and {Josaitis}, Alec and {Kern}, Nicholas S. and {Kerrigan}, Joshua and {Kittiwisit}, Piyanat and {Kolopanis}, Matthew and {Plante}, Paul La and {Lanman}, Adam and {Ma}, Yin-Zhe and {MacMahon}, David H.~E. and {Malan}, Lourence and {Malgas}, Cresshim and {Malgas}, Keith and {Marero}, Bradley and {Martinot}, Zachary E. and {McBride}, Lisa and {Mesinger}, Andrei and {Mohamed-Hinds}, Nicel and {Molewa}, Mathakane and {Morales}, Miguel F. and {Murray}, Steven G. and {Nuwegeld}, Hans and {Parsons}, Aaron R. and {Pascua}, Robert and {Qin}, Yuxiang and {Rath}, Eleanor and {Razavi-Ghods}, Nima and {Robnett}, James and {Santos}, Mario G. and {Sims}, Peter and {Singh}, Saurabh and {Storer}, Dara and {Swarts}, Hilton and {Tan}, Jianrong and {Wyngaarden}, Pieter van and {Zheng}, Haoxuan},
        title = "{Impacts and Statistical Mitigation of Missing Data on the 21 cm Power Spectrum: A Case Study with the Hydrogen Epoch of Reionization Array}",
      journal = {\apj},
         year = 2025,
        month = feb,
       volume = {979},
       number = {2},
          eid = {191},
        pages = {191},
          doi = {10.3847/1538-4357/ad9b91},
archivePrefix = {arXiv},
       eprint = {2411.10529},
 primaryClass = {astro-ph.CO},
       adsurl = {https://ui.adsabs.harvard.edu/abs/2025ApJ...979..191C}
}

@INPROCEEDINGS{CMB_2022,
       author = {{Barron}, Darcy R. and {Bender}, Amy N. and {Birdwell}, Ian E. and {Carlstrom}, John E. and {Delabrouille}, Jacques and {Guns}, Sam and {Kovac}, John and {Lawrence}, Charles R. and {Paine}, Scott and {Whitehorn}, Nathan},
        title = "{Review of radio frequency interference and potential impacts on the CMB-S4 cosmic microwave background survey}",
    booktitle = {Millimeter, Submillimeter, and Far-Infrared Detectors and Instrumentation for Astronomy XI},
         year = 2022,
       editor = {{Zmuidzinas}, Jonas and {Gao}, Jian-Rong},
       series = {Society of Photo-Optical Instrumentation Engineers (SPIE) Conference Series},
       volume = {12190},
        month = aug,
          eid = {1219002},
        pages = {1219002},
          doi = {10.1117/12.2629651},
archivePrefix = {arXiv},
       eprint = {2207.13204},
 primaryClass = {astro-ph.IM},
       adsurl = {https://ui.adsabs.harvard.edu/abs/2022SPIE12190E..02B}
}

@ARTICLE{Offringa_2012,
       author = {{Offringa}, A.~R. and {van de Gronde}, J.~J. and {Roerdink}, J.~B.~T.~M.},
        title = "{A morphological algorithm for improving radio-frequency interference detection}",
      journal = {\aap},
         year = 2012,
        month = mar,
       volume = {539},
          eid = {A95},
        pages = {A95},
          doi = {10.1051/0004-6361/201118497},
archivePrefix = {arXiv},
       eprint = {1201.3364},
 primaryClass = {astro-ph.IM},
       adsurl = {https://ui.adsabs.harvard.edu/abs/2012A&A...539A..95O}
}

@ARTICLE{Ben_david_2008,
       author = {{Ben-David}, Chen and {Leshem}, Amir},
        title = "{Parametric High Resolution Techniques for Radio Astronomical Imaging}",
      journal = {IEEE Journal of Selected Topics in Signal Processing},
         year = 2008,
        month = oct,
       volume = {2},
       number = {5},
        pages = {670-684},
          doi = {10.1109/JSTSP.2008.2005318},
archivePrefix = {arXiv},
       eprint = {0807.4852},
 primaryClass = {astro-ph},
       adsurl = {https://ui.adsabs.harvard.edu/abs/2008ISTSP...2..670B}
}

@INPROCEEDINGS{van_der_veen_2004,
  author={van der Veen, A.-J. and Amir Leshem and Boonstra, A.-J.},
  booktitle={Processing Workshop Proceedings, 2004 Sensor Array and Multichannel Signal}, 
  title={Signal processing for radio astronomical arrays}, 
  year={2004},
  volume={},
  number={},
  pages={1-10},
  doi={10.1109/SAM.2004.1502901}}

@ARTICLE{Hellbourg_2018,
       author = {{Hellbourg}, Gregory},
        title = "{RFI subspace smearing and projection for array radio telescopes}",
      journal = {arXiv e-prints},
         year = 2018,
        month = sep,
          eid = {arXiv:1809.03620},
        pages = {arXiv:1809.03620},
          doi = {10.48550/arXiv.1809.03620},
archivePrefix = {arXiv},
       eprint = {1809.03620},
 primaryClass = {eess.SP},
       adsurl = {https://ui.adsabs.harvard.edu/abs/2018arXiv180903620H}
}

@ARTICLE{Masui_2017,
       author = {{Masui}, Kiyoshi W. and {Shaw}, J. Richard and {Ng}, Cherry and {Smith}, Kendrick M. and {Vanderlinde}, Keith and {Paradise}, Adiv},
        title = "{Algorithms for FFT Beamforming Radio Interferometers}",
      journal = {\apj},
         year = 2019,
        month = jul,
       volume = {879},
       number = {1},
          eid = {16},
        pages = {16},
          doi = {10.3847/1538-4357/ab229e},
archivePrefix = {arXiv},
       eprint = {1710.08591},
 primaryClass = {astro-ph.IM},
       adsurl = {https://ui.adsabs.harvard.edu/abs/2019ApJ...879...16M}
}

@ARTICLE{Capon_1969,
  author={Capon, J.},
  journal={Proceedings of the IEEE}, 
  title={High-resolution frequency-wavenumber spectrum analysis}, 
  year={1969},
  volume={57},
  number={8},
  pages={1408-1418},
  doi={10.1109/PROC.1969.7278}}

@ARTICLE{Michilli_2021,
       author = {{Michilli}, D. and {Masui}, K.~W. and {Mckinven}, R. and {Cubranic}, D. and {Bruneault}, M. and {Brar}, C. and {Patel}, C. and {Boyle}, P.~J. and {Stairs}, I.~H. and {Renard}, A. and {Bandura}, K. and {Berger}, S. and {Breitman}, D. and {Cassanelli}, T. and {Dobbs}, M. and {Kaspi}, V.~M. and {Leung}, C. and {Mena-Parra}, J. and {Pleunis}, Z. and {Russell}, L. and {Scholz}, P. and {Siegel}, S.~R. and {Tendulkar}, S.~P. and {Vanderlinde}, K.},
        title = "{An Analysis Pipeline for CHIME/FRB Full-array Baseband Data}",
      journal = {\apj},
         year = 2021,
        month = apr,
       volume = {910},
       number = {2},
          eid = {147},
        pages = {147},
          doi = {10.3847/1538-4357/abe626},
archivePrefix = {arXiv},
       eprint = {2010.06748},
 primaryClass = {astro-ph.HE},
       adsurl = {https://ui.adsabs.harvard.edu/abs/2021ApJ...910..147M}
}

@preamble{ " \newcommand{\noop}[1]{} " }

@ARTICLE{chime_overview,
       author = {{CHIME Collaboration} and {Amiri}, Mandana and {Bandura}, Kevin and {Boskovic}, Anja and {Chen}, Tianyue and {Cliche}, Jean-Fran{\c{c}}ois and {Deng}, Meiling and {Denman}, Nolan and {Dobbs}, Matt and {Fandino}, Mateus and {Foreman}, Simon and {Halpern}, Mark and {Hanna}, David and {Hill}, Alex S. and {Hinshaw}, Gary and {H{\"o}fer}, Carolin and {Kania}, Joseph and {Klages}, Peter and {Landecker}, T.~L. and {MacEachern}, Joshua and {Masui}, Kiyoshi and {Mena-Parra}, Juan and {Milutinovic}, Nikola and {Mirhosseini}, Arash and {Newburgh}, Laura and {Nitsche}, Rick and {Ordog}, Anna and {Pen}, Ue-Li and {Pinsonneault-Marotte}, Tristan and {Polzin}, Ava and {Reda}, Alex and {Renard}, Andre and {Shaw}, J. Richard and {Siegel}, Seth R. and {Singh}, Saurabh and {Smegal}, Rick and {Tretyakov}, Ian and {van Gassen}, Kwinten and {Vanderlinde}, Keith and {Wang}, Haochen and {Wiebe}, Donald V. and {Willis}, James S. and {Wulf}, Dallas},
        title = "{An Overview of CHIME, the Canadian Hydrogen Intensity Mapping Experiment}",
      journal = {\apjs},
         year = 2022,
        month = aug,
       volume = {261},
       number = {2},
          eid = {29},
        pages = {29},
          doi = {10.3847/1538-4365/ac6fd9},
archivePrefix = {arXiv},
       eprint = {2201.07869},
 primaryClass = {astro-ph.IM},
       adsurl = {https://ui.adsabs.harvard.edu/abs/2022ApJS..261...29C}
}

@ARTICLE{kko_adam,
       author = {{Lanman}, Adam E. and {Andrew}, Shion and {Lazda}, Mattias and {Shah}, Vishwangi and {Amiri}, Mandana and {Balasubramanian}, Arvind and {Bandura}, Kevin and {Boyle}, P.~J. and {Brar}, Charanjot and {Carlson}, Mark and {Cliche}, Jean-Fran{\c{c}}ois and {Gusinskaia}, Nina and {Hendricksen}, Ian T. and {Kaczmarek}, J.~F. and {Landecker}, Tom and {Leung}, Calvin and {Mckinven}, Ryan and {Mena-Parra}, Juan and {Milutinovic}, Nikola and {Nimmo}, Kenzie and {Pearlman}, Aaron B. and {Renard}, Andre and {Rahman}, Mubdi and {Shaw}, J. Richard and {Siegel}, Seth R. and {Smegal}, Rick J. and {Cassanelli}, Tomas and {Chatterjee}, Shami and {Curtin}, Alice P. and {Dobbs}, Matt and {Dong}, Fengqiu Adam and {Halpern}, Mark and {Hopkins}, Hans and {Kaspi}, Victoria M. and {Khairy}, Kholoud and {Masui}, Kiyoshi W. and {Meyers}, Bradley W. and {Michilli}, Daniele and {Petroff}, Emily and {Pinsonneault-Marotte}, Tristan and {Pleunis}, Ziggy and {Rafiei-Ravandi}, Masoud and {Shin}, Kaitlyn and {Smith}, Kendrick and {Vanderlinde}, Keith and {Zegmott}, Tarik J.},
        title = "{CHIME/FRB Outriggers: KKO Station System and Commissioning Results}",
      journal = {\aj},
         year = 2024,
        month = aug,
       volume = {168},
       number = {2},
          eid = {87},
        pages = {87},
          doi = {10.3847/1538-3881/ad5838},
archivePrefix = {arXiv},
       eprint = {2402.07898},
 primaryClass = {astro-ph.IM},
       adsurl = {https://ui.adsabs.harvard.edu/abs/2024AJ....168...87L}
}

@misc{menaparra_quantization_bias,
      title={Quantization bias for digital correlators}, 
      author={J. Mena-Parra and K. Bandura and M. A. Dobbs and J. R. Shaw and S. Siegel},
      year={2018},
      eprint={1803.04296},
      archivePrefix={arXiv},
      primaryClass={astro-ph.IM},
      url={https://arxiv.org/abs/1803.04296}, 
}

@ARTICLE{chime_frb_2018,
       author = {{CHIME/FRB Collaboration} and {Amiri}, M. and {Bandura}, K. and {Berger}, P. and {Bhardwaj}, M. and {Boyce}, M.~M. and {Boyle}, P.~J. and {Brar}, C. and {Burhanpurkar}, M. and {Chawla}, P. and {Chowdhury}, J. and {Cliche}, J. -F. and {Cranmer}, M.~D. and {Cubranic}, D. and {Deng}, M. and {Denman}, N. and {Dobbs}, M. and {Fandino}, M. and {Fonseca}, E. and {Gaensler}, B.~M. and {Giri}, U. and {Gilbert}, A.~J. and {Good}, D.~C. and {Guliani}, S. and {Halpern}, M. and {Hinshaw}, G. and {H{\"o}fer}, C. and {Josephy}, A. and {Kaspi}, V.~M. and {Landecker}, T.~L. and {Lang}, D. and {Liao}, H. and {Masui}, K.~W. and {Mena-Parra}, J. and {Naidu}, A. and {Newburgh}, L.~B. and {Ng}, C. and {Patel}, C. and {Pen}, U. -L. and {Pinsonneault-Marotte}, T. and {Pleunis}, Z. and {Rafiei Ravandi}, M. and {Ransom}, S.~M. and {Renard}, A. and {Scholz}, P. and {Sigurdson}, K. and {Siegel}, S.~R. and {Smith}, K.~M. and {Stairs}, I.~H. and {Tendulkar}, S.~P. and {Vanderlinde}, K. and {Wiebe}, D.~V.},
        title = "{The CHIME Fast Radio Burst Project: System Overview}",
      journal = {\apj},
         year = 2018,
        month = aug,
       volume = {863},
       number = {1},
          eid = {48},
        pages = {48},
          doi = {10.3847/1538-4357/aad188},
archivePrefix = {arXiv},
       eprint = {1803.11235},
 primaryClass = {astro-ph.IM},
       adsurl = {https://ui.adsabs.harvard.edu/abs/2018ApJ...863...48C}
}

@ARTICLE{leung2020synoptic,
       author = {{Leung}, Calvin and {Mena-Parra}, Juan and {Masui}, Kiyoshi and {Bandura}, Kevin and {Bhardwaj}, Mohit and {Boyle}, P.~J. and {Brar}, Charanjot and {Bruneault}, Mathieu and {Cassanelli}, Tomas and {Cubranic}, Davor and {Kaczmarek}, Jane F. and {Kaspi}, Victoria and {Landecker}, Tom and {Michilli}, Daniele and {Milutinovic}, Nikola and {Patel}, Chitrang and {Pleunis}, Ziggy and {Rahman}, Mubdi and {Renard}, Andre and {Sanghavi}, Pranav and {Stairs}, Ingrid H. and {Scholz}, Paul and {Vanderlinde}, Keith and {Chime/Frb Collaboration}},
        title = "{A Synoptic VLBI Technique for Localizing Nonrepeating Fast Radio Bursts with CHIME/FRB}",
      journal = {\aj},
         year = 2021,
        month = feb,
       volume = {161},
       number = {2},
          eid = {81},
        pages = {81},
          doi = {10.3847/1538-3881/abd174},
archivePrefix = {arXiv},
       eprint = {2008.11738},
 primaryClass = {astro-ph.IM},
       adsurl = {https://ui.adsabs.harvard.edu/abs/2021AJ....161...81L}
}

@INPROCEEDINGS{Deller_2010,
       author = {{Deller}, A.},
        title = "{Software correlators as testbeds for RFI algorithms}",
    booktitle = {RFI Mitigation Workshop},
         year = 2010,
        month = may,
          eid = {35},
        pages = {35},
          doi = {10.22323/1.107.0035},
archivePrefix = {arXiv},
       eprint = {1012.0325},
 primaryClass = {astro-ph.IM},
       adsurl = {https://ui.adsabs.harvard.edu/abs/2010rfim.workE..35D}
}

@ARTICLE{RBFLOAT_2025ApJ,
       author = {{CHIME/FRB Collaboration} and {Abbott}, Thomas C. and {Amouyal}, Daniel and {Andersen}, Bridget C. and {Andrew}, Shion E. and {Bandura}, Kevin and {Bhardwaj}, Mohit and {Bhopi}, Kalyani and {Bhusare}, Yash and {Brar}, Charanjot and {Cai}, Alice and {Cassanelli}, Tomas and {Chatterjee}, Shami and {Cliche}, Jean-Fran{\c{c}}ois and {Cook}, Amanda M. and {Curtin}, Alice P. and {Davies-Velie}, Evan and {Dobbs}, Matt and {Dong}, Fengqiu Adam and {Dong}, Yuxin and {Eadie}, Gwendolyn and {Eftekhari}, Tarraneh and {Fong}, Wen-Fai and {Fonseca}, Emmanuel and {Gaensler}, B.~M. and {Gusinskaia}, Nina and {Hessels}, Jason W.~T. and {Hewitt}, Dant{\'e} M. and {Huang}, Jeff and {Jain}, Naman and {Joseph}, Ronniy. C. and {Kahinga}, Lordrick and {Kaspi}, Victoria M. and {Khan}, Afrasiyab (Afrokk) and {Kharel}, Bikash and {Lanman}, Adam E. and {L'Argent}, Magnus and {Lazda}, Mattias and {Leung}, Calvin and {Main}, Robert and {Mas-Ribas}, Lluis and {Masui}, Kiyoshi W. and {McGregor}, Kyle and {McKinven}, Ryan and {Mena-Parra}, Juan and {Michilli}, Daniele and {Mulyk}, Nicole and {Ng}, Mason and {Nimmo}, Kenzie and {Pandhi}, Ayush and {Patil}, Swarali Shivraj and {Pearlman}, Aaron B. and {Pen}, Ue-Li and {Pleunis}, Ziggy and {Prochaska}, J. Xavier and {Rafiei-Ravandi}, Masoud and {Ransom}, Scott M. and {Sachdeva}, Gurman and {Sammons}, Mawson W. and {Sand}, Ketan R. and {Scholz}, Paul and {Shah}, Vishwangi and {Shin}, Kaitlyn and {Siegel}, Seth R. and {Simha}, Sunil and {Smith}, Kendrick and {Stairs}, Ingrid and {Stenning}, David C. and {Wang}, Haochen and {Boles}, Thomas and {Cognard}, Isma{\"e}l and {Dijkema}, Tammo Jan and {Filippenko}, Alexei V. and {Gawro{\'n}ski}, Marcin P. and {Herrmann}, Wolfgang and {Kilpatrick}, Charles D. and {Kirsten}, Franz and {Knabel}, Shawn and {Ould-Boukattine}, Omar S. and {Paugnat}, Hadrien and {Puchalska}, Weronika and {Sheu}, William and {Suresh}, Aswin and {Tohuvavohu}, Aaron and {Treu}, Tommaso and {Zheng}, Weikang},
        title = "{FRB 20250316A: A Brilliant and Nearby One-off Fast Radio Burst Localized to 13 pc Precision}",
      journal = {\apjl},
         year = 2025,
        month = aug,
       volume = {989},
       number = {2},
          eid = {L48},
        pages = {L48},
          doi = {10.3847/2041-8213/adf62f},
archivePrefix = {arXiv},
       eprint = {2506.19006},
 primaryClass = {astro-ph.HE},
       adsurl = {https://ui.adsabs.harvard.edu/abs/2025ApJ...989L..48C}
}

@ARTICLE{Andrew_2025,
       author = {{Andrew}, Shion and {Leung}, Calvin and {Li}, Alexander and {Masui}, Kiyoshi W. and {Andersen}, Bridget C. and {Bandura}, Kevin and {Curtin}, Alice P. and {Kaczmarek}, Jane and {Lanman}, Adam E. and {Lazda}, Mattias and {Mena-Parra}, Juan and {Michilli}, Daniele and {Nimmo}, Kenzie and {Pearlman}, Aaron B. and {Rahman}, Mubdi and {Shah}, Vishwangi and {Shin}, Kaitlyn and {Wang}, Haochen and {Chime/Frb Collaboration}},
        title = "{A Very Long Baseline Interferometry Calibrator Grid at 600 MHz for Fast Radio Transient Localizations with CHIME/FRB Outriggers}",
      journal = {\apj},
         year = 2025,
        month = mar,
       volume = {981},
       number = {1},
          eid = {39},
        pages = {39},
          doi = {10.3847/1538-4357/adaf8d},
archivePrefix = {arXiv},
       eprint = {2409.11476},
 primaryClass = {astro-ph.IM},
       adsurl = {https://ui.adsabs.harvard.edu/abs/2025ApJ...981...39A}
}

@ARTICLE{Leung_2025,
       author = {{Leung}, Calvin and {Andrew}, Shion and {Masui}, Kiyoshi W. and {Brar}, Charanjot and {Cassanelli}, Tomas and {Chatterjee}, Shami and {Kaspi}, Victoria and {Khairy}, Kholoud and {Lanman}, Adam E. and {Lazda}, Mattias and {Mena-Parra}, Juan and {Noble}, Gavin and {Pearlman}, Aaron B. and {Rahman}, Mubdi and {Sanghavi}, Pranav and {Shah}, Vishwangi and {Chime/Frb Collaboration}},
        title = "{A VLBI Software Correlator for Fast Radio Transients}",
      journal = {\aj},
         year = 2025,
        month = jul,
       volume = {170},
       number = {1},
          eid = {53},
        pages = {53},
          doi = {10.3847/1538-3881/add876},
archivePrefix = {arXiv},
       eprint = {2403.05631},
 primaryClass = {astro-ph.IM},
       adsurl = {https://ui.adsabs.harvard.edu/abs/2025AJ....170...53L}
}

@INPROCEEDINGS{Hallinan_2019,
       author = {{Hallinan}, Gregg and {Ravi}, V. and {Weinreb}, S. and {Kocz}, J. and {Huang}, Y. and {Woody}, D.~P. and {Lamb}, J. and {D'Addario}, L. and {Catha}, M. and {Law}, C. and {Kulkarni}, S.~R. and {Phinney}, E.~S. and {Eastwood}, M.~W. and {Bouman}, K. and {McLaughlin}, M. and {Ransom}, S. and {Siemens}, X. and {Cordes}, J. and {Lynch}, R. and {Kaplan}, D. and {Brazier}, A. and {Bhatnagar}, S. and {Myers}, S. and {Walter}, F. and {Gaensler}, B.},
        title = "{The DSA-2000 {\textemdash} A Radio Survey Camera}",
    booktitle = {Bulletin of the American Astronomical Society},
         year = 2019,
       volume = {51},
        month = sep,
          eid = {255},
        pages = {255},
          doi = {10.48550/arXiv.1907.07648},
archivePrefix = {arXiv},
       eprint = {1907.07648},
 primaryClass = {astro-ph.IM},
       adsurl = {https://ui.adsabs.harvard.edu/abs/2019BAAS...51g.255H}
}

@book{Interferometry_Thompson,
  author    = {{Thompson}, A.~R. and {Moran}, J.~M. and {Swenson Jr}, G.~W.}, 
  title     = {Interferometry and Synthesis in Radio Astronomy},
  publisher = {Springer, Cham},
  year      = 2017,
  edition   = 3,
  isbn      = {978-3-319-44431-4}
}

@ARTICLE{2015_Masui,
       author = {{Masui}, K. and {Amiri}, M. and {Connor}, L. and {Deng}, M. and {Fandino}, M. and {H{\"o}fer}, C. and {Halpern}, M. and {Hanna}, D. and {Hincks}, A.~D. and {Hinshaw}, G. and {Parra}, J.~M. and {Newburgh}, L.~B. and {Shaw}, J.~R. and {Vanderlinde}, K.},
        title = "{A compression scheme for radio data in high performance computing}",
      journal = {Astronomy and Computing},
         year = 2015,
        month = sep,
       volume = {12},
        pages = {181-190},
          doi = {10.1016/j.ascom.2015.07.002},
archivePrefix = {arXiv},
       eprint = {1503.00638},
 primaryClass = {astro-ph.IM},
       adsurl = {https://ui.adsabs.harvard.edu/abs/2015A&C....12..181M}
}

@article{FERRANDI2025108108,
title = {A subspace method for large-scale trace ratio problems},
journal = {Computational Statistics \& Data Analysis},
volume = {205},
pages = {108108},
year = {2025},
issn = {0167-9473},
doi = {https://doi.org/10.1016/j.csda.2024.108108},
url ={https://www.sciencedirect.com/science/article/pii/S0167947324001920},
author = {Giulia Ferrandi and Michiel E. Hochstenbach and M. {Rosário Oliveira}},
}

@article{doi:10.1137/120864799,
author = {Ngo, T. T. and Bellalij, M. and Saad, Y.},
title = {The Trace Ratio Optimization Problem},
journal = {SIAM Review},
volume = {54},
number = {3},
pages = {545-569},
year = {2012},
doi = {10.1137/120864799},
URL = {https://doi.org/10.1137/120864799},
eprint = {https://doi.org/10.1137/120864799}
}

@INPROCEEDINGS{2025_CHARTS,
  author={Cassanelli, Tomás and Mena-Parra, Juan and Manosalva, Sebastián and Lau, Albert Wai Kit and Gallardo, Diego and Vanderlinde, Keith and Connor, Liam and Finger, Ricardo and Hum, Sean},
  booktitle={2025 19th European Conference on Antennas and Propagation (EuCAP)}, 
  title={Canadian-Chilean Array for Radio Transient Studies (CHARTS): Analog System Developments}, 
  year={2025},
  volume={},
  number={},
  pages={1-5},
  doi={10.23919/EuCAP63536.2025.10999353}}

\bibliographystyle{aasjournal}
\end{document}